\documentclass{LMCS} 

\usepackage{latexsym}
\usepackage{amsmath,amsthm}
\usepackage{amssymb}
\usepackage{pstricks}
\usepackage{pst-node}
\usepackage{prettyref}
\newrefformat{lem}{Lemma~\ref{#1}}
\newrefformat{the}{Theorem~\ref{#1}}
\newrefformat{cor}{Corollary~\ref{#1}}
\newrefformat{alg}{Algorithm~\ref{#1}}
\newrefformat{pro}{Proposition~\ref{#1}}
\newrefformat{fig}{Figure~\ref{#1}}
\newrefformat{cha}{Chapter~\ref{#1}}
\newrefformat{sec}{Section~\ref{#1}}
\newrefformat{eqn}{(\ref{#1})}

\newcommand{\pref}[1]{\prettyref{#1}}

\newcommand{\rest}[2]{\ensuremath{#1|_{#2}}}

\newcommand{\var}[0]{\ensuremath{\mathit{Var}}}
\newcommand{\dom}[0]{\ensuremath{\mathit{dom}}}
\newcommand{\set}[2]{ \{$ $#1$ $|$ $#2$ $\} }

\newcommand{\ev}[1]{\ensuremath{\mathit{EVar}(#1)}}

\newcommand{\ve}[1]{\ensuremath{\mathit{VEx}(#1)}}

\newcommand{\wres}[0]{\ensuremath{\vdash^w}}

\def\DLLlearn{{\textsc{DLL-Learn}}}

\newcommand{\graph}[2]{\ensuremath{{#2}_{#1}}}

\newcommand{\leafs}[1]{\ensuremath{V_{#1}^0}}

\newcommand{\Cc}[1]{\ensuremath{CC({#1})}}
\newcommand{\CC}[1]{\ensuremath{CC(#1)}}

\newcommand{\ic}[2]{\ensuremath{IC({#1},{#2})}}

\usepackage{enumerate}%,hyperref}

\def\doi{4 (4:13) 2008}
\lmcsheading%
{\doi}
{1--28}
{}
{}
{Jun.~24, 2008}
{Dec.~\phantom{0}5, 2008}
{}   

\begin{document}

\title[Resolution Trees with Lemmas]{Resolution Trees with Lemmas: \\
  Resolution Refinements that Characterize DLL Algorithms with Clause
  Learning}

\author[S.~R.~Buss]{Samuel R.\ Buss\rsuper a}
\address{{\lsuper a}Department of Mathematics \\
         University of California, San Diego\\
         La Jolla, CA 92093-0112, USA}
\email{sbuss@math.ucsd.edu}
\thanks{{\lsuper a}Supported in part by NSF grants DMS-0400848 and DMS-0700533.}

\author[J.~Hoffmann]{Jan Hoffmann\rsuper b}
\address{{\lsuper{b,c}}Institut f\"ur Informatik\\
         Ludwig-Maximilians Universit\"at\\
         D-80538 M\"unchen, Germany}
\email{\{jan.hoffmann,jan.johannsen\}@ifi.lmu.de}
\thanks{{\lsuper b}Supported in part by the
Studienstiftung des deutschen Volkes (German National Merit
Foundation).}

\author[J.~Johannsen]{Jan Johannsen\rsuper c}
%\address{Institut f\"ur Informatik\\
%         Ludwig-Maximilians Universit\"at\\
%         D-80538 M\"unchen, Germany}
%\email{jan.johannsen@ifi.lmu.de}

%% required for running head on odd and even pages, use suitable
%% abbreviations in case of long titles and many authors:

%% mandatory lists of keywords and classifications:
\keywords{propositional proof complexity, resolution, SAT solving, DLL
  algoritm, clause learning}
\subjclass{F.2.2, I.2.8}

% \titlecomment{OPTIONAL comment concerning the title, \eg, if a variant
% or an extended abstract of the paper has appeared elsewehere}

\begin{abstract}
  Resolution refinements called \emph{w-resolution trees with
    lemmas} (WRTL) and with \emph{input lemmas} (WRTI) are introduced.
  Dag-like resolution is equivalent to both WRTL and WRTI when there
  is no regularity condition.  For regular proofs, an exponential
  separation between regular dag-like resolution and both regular WRTL
  and regular WRTI is given.

  It is proved that DLL proof search algorithms that use clause
  learning based on unit propagation can be polynomially simulated by
  regular WRTI.  More generally, non-greedy DLL algorithms with
  learning by unit propagation are equivalent to regular WRTI.  A
  general form of clause learning, called DLL-Learn, is defined that
  is equivalent to regular WRTL.

  A variable extension method is used to give simulations of
  resolution by regular WRTI, using a simplified form of proof trace
  extensions.  DLL-Learn and non-greedy DLL algorithms with learning
  by unit propagation can use variable extensions to simulate general
  resolution without doing restarts.

  Finally, an exponential lower bound for WRTL where the lemmas are
  restricted to short clauses is shown.
\end{abstract}

\maketitle

\section{Introduction}\label{sec:intro}

Although
the satisfiability problem for propositional logic (SAT) is NP-complete,
there exist SAT solvers that can decide SAT on present-day computers
for many formulas that are relevant in practice
\cite{SilvaSakallah1996, MoskewiczMalik2001, MahajanFu2004,
  BerreSimon2003, BerreSimon2004, BerreSimon2005}.  
The fastest SAT solvers for structured problems are based on the basic
backtracking procedures known as DLL algorithms
\cite{DavisLogemann1962}, extended with additional techniques such as
clause learning.   
 
DLL algorithms can be seen as a kind of proof search procedure since
the execution of a DLL algorithm on an unsatisfiable CNF formula
yields a tree-like resolution refutation of that formula.  Conversely,
given a tree-like resolution refutation, an execution of a DLL
algorithm on the refuted formula can be constructed whose runtime is
roughly the size of the refutation. By this exact correspondence,
upper and lower bounds on the size of tree-like resolution proofs
transfer to bounds on the runtime of DLL algorithms.

This paper generalizes this exact correspondence to extensions of DLL
by clause learning.  To this end, we define natural, rule-based
resolution proof systems and then prove that they correspond to DLL
algorithms that use various forms of clause learning. 
The motivation for this is that
the correspondence between
a clause learning DLL algorithm and a proof system helps explain the 
power of the algorithm by giving
a description of the space of proofs which is
searched by it.  
In addition, upper and lower bounds
on proof complexity can be transferred to upper and lower bounds on the
possible runtimes of large classes of DLL algorithms with clause learning.

% By experience,
% complexity results such as simulations, separations and lower bounds
% are easier to prove for proof systems than for families of algorithms.
% The last two sections give expamles of such results for DLL algorithms
% with clause learning that are shown for the corresponding proof
% systems.  An exact description of the set of resolution proofs that is
% searched by clause learning DLL algorithms also provides alternative
% ways to describe and understand these algorithms.
% 
% 
% This clear distinction between algorithms and proof systems is in
% contrast to other works relating clause learning algorithms to
% resolution proof systems by Beame et al.
% \cite{BeameKautzSabharwal2004}, Van Gelder \cite{VanGelder2005} and
% Bacchus et al. \cite{BHPvG:clauselearn}, who define proof systems in
% terms of algorithms, and then compare these to other known proof
% systems.
% 

We introduce, in {\pref{sec:wrtl}}, tree-like
resolution refinements using the notions of a resolution tree with
lemmas (RTL) and a resolution tree with input lemmas (RTI). 
An RTL is a tree-like resolution proof in which every
clause needs only to be derived once and can be copied to be used as a
leaf in the tree (i.e., a lemma) if it is used several times.  As the
reader might guess, RTL is polynomially equivalent to general
resolution.

Since DLL algorithms use learning based on unit propagation, and since
unit propagation is equivalent to input resolution (sometimes
called ``trivial resolution'' \cite{BeameKautzSabharwal2004}), it is
useful to restrict the lemmas that are used in a RTL to those that
appear as the root of input subproofs.
This gives rise to proof systems based on resolution
trees with input lemmas (RTI).  Somewhat surprisingly, we
show that RTI can also simulate general resolution.

A resolution proof is called {\em regular} if no variable is used as a
resolution variable twice along any path in the tree. Regular proofs
occur naturally in the present context, since a backtracking algorithm
would never query the same variable twice on one branch of its
execution.
It is known that regular resolution is weaker than
general resolution~\cite{Goerdt1993,Alekhnovich2002},
but it is unknown whether
regular resolution can simulate regular RTL or regular RTI.  This
is because, in regular RTL/RTI proofs, variables that are used for
resolution to derive a clause can be reused on paths where this clause
appears as a lemma.

For resolution and regular resolution,
the use of a weakening rule does not increase the power
of the proof system (by the subsumption principle).  However,
for RTI and regular RTL proofs, the weakening rule may increase the strength
of the proof system (this is an open question, in fact),
since eliminating uses
of weak inferences may require pruning away parts of the proof that contain
lemmas needed later in the proof.
Accordingly, \pref{sec:wrtl} also defines proof systems
regWRTL and regWRTI that consist of regular RTL and regular RTI
(respectively), but with a modified form
of resolution, called ``w-resolution'', that incorporates
a restricted form
of the weakening rule.

In {\pref{sec:dll-up}} we propose a general framework
for DLL
algorithms with clause learning, called \textsc{DLL-L-UP}. 
The schema \textsc{DLL-L-UP} is an attempt to give a short and abstract
definition of 
modern SAT solvers and it
incorporates all common learning
strategies, including all the specific strategies discussed by
Beame et al.~\cite{BeameKautzSabharwal2004}.
{\pref{sec:regwrti-dll}} proves that,
for any of these learning strategies, a proof search tree
can be transformed into a regular WRTI proof with only a polynomial
increase in size.  Conversely, any regular WRTI proof can be simulated by 
a ``non-greedy'' DLL search tree with clause learning, where by ``non-greedy''
is meant that the algorithm can continue decision branching even after
unit propagation could yield a contradiction. 

In {\pref{sec:dll-learn}} we give another
generalization of
DLL with clause learning called \DLLlearn.
The algorithm \DLLlearn{} can simulate the clause learning algorithm
\textsc{DLL-L-UP}.
More precisely, we prove that
\DLLlearn{} p-simulates, and is 
p-simulated by, regular WRTL.
The
\DLLlearn{} algorithm
is very similar to the ``pool resolution'' algorithm that
has been introduced by Van Gelder~\cite{VanGelder2005}
but differs from pool resolution by using the
``w-resolution'' inference in place of the ``degenerate'' inference
used by Van Gelder (the terminology ``degenerate'' is
used by Hertel et al.~\cite{BHPvG:clauselearn}).  
Van Gelder has shown that
pool resolution can simulate not only regular resolution, but
also any resolution refutation which has a regular depth-first search
tree.
The latter proof system is the same as
the proof system regRTL in our framework, therefore 
the same holds for \DLLlearn{}.
It is unknown whether \DLLlearn{} or \textsc{DLL-L-UP} can p-simulate
pool resolution or vice versa.

Sections \ref{sec:dll-up}-\ref{sec:dll-learn} prove the equivalence 
of clause learning algorithms with the two proof systems regWRTI and
regWRTL.  
Our really novel system is regWRTI: this system has the 
advantage of using
input lemmas in a manner that closely matches the range of clause
learning algorithms that can be used by practical DLL algorithms.
In particular, the regWRTI proof system's use of input lemmas
corresponds directly to the clause learning strategies
of Silva and Sakallah \cite{SilvaSakallah1996}, including
first-UIP, relsat, and other clauses based on cuts, and
including learning multiple clauses at a time.  
Van Gelder~\cite{VanGelder2005} shows that pool resolution can also
simulate these kinds of clause learning (at least, for learning
single clauses), but the correspondence is much
more natural for the system regWRTI than for either pool resolution
or \DLLlearn{}.

It is known that DLL algorithms with clause learning
and restarts can simulate full (non-regular, dag-like)
resolution by learning every derived clause,
and doing a restart
each time a clause is learned~\cite{BeameKautzSabharwal2004}. 
Our proof systems, regWRTI and \DLLlearn{}, do not handle 
restarts; instead, they can be viewed as capturing what can happen
between restarts.  Another approach to simulating full resolution
is via the use of ``proof trace extensions'' introduced by 
Beame et al.~\cite{BeameKautzSabharwal2004}.  
Proof trace extensions allow resolution to be simulated by
clause learning DLL algorithms, and a related construction is
used by Hertel et al.~\cite{BHPvG:clauselearn}
to show that pool resolution can ``effectively'' 
p-simulate full resolution.   These constructions require introducing
new variables and clauses in a way that does not affect satisfiability, but
allow a clause learning
DLL algorithm or pool resolution to establish non-satisfiability.
However,
the constructions by Beame et al.~\cite{BeameKautzSabharwal2004} and the initially
circulated
preprint of Hertel et al.~\cite{BHPvG:clauselearn} had
the drawback that the number of extra
introduced variables depends on the size of the (unknown) resolution
refutation.  

\pref{sec:varexp} introduces an improved form of proof trace
extensions called
``variable extensions''.  Theorem~\ref{the:pte_trick} shows that
variable extensions can be used to give a p-simulation
of full resolution by regWRTI (at the cost of changing the
formula that is being refuted).  
Variable extensions are simpler and more powerful than proof
trace extensions.
Their main advantage 
is that a variable extension depends only on the number of variables,
not on the size of the (unknown) resolution proof.
The results of \pref{sec:varexp} were first published 
in the second author's diploma thesis \cite{Hoffmann2007};
the subsequently published version of the article of Hertel et al.~\cite{BHPvG:clauselearn}
gives a similarly improved construction (for pool resolution)
that does not depend on the
size of the resolution proof and, in addition, does not use
degenerate resolution inferences.

One consequence of Theorem~\ref{the:pte_trick} is that
regWRTI can effectively p-simulate full resolution.  This 
improves on the results of Hertel et al.~\cite{BHPvG:clauselearn} 
since regWRTI is not known to be
as strong as pool resolution.
It remains open whether regWRTI or pool resolution
can p-simulate general resolution without variable extensions.

\pref{sec:smlem}
proves a lower bound that shows that for certain hard formulas, 
the pigeonhole principle $PHP_n$, learning only small
clauses does not help a DLL-algorithm.  We show that resolution trees
with lemmas require size exponential in $n\log n$ to refute $PHP_n$
when the size of clauses used as lemmas is restricted to be less than
$n/2$.  This bound is asymptotically the same as the lower bound shown
for tree-like resolution refutations of $PHP_n$ \cite{iwamiy99}.  On
the other hand, there are
regular resolution refutations of $PHP_n$
of size exponential in~$n$~\cite{BusPit97},
and our results show that
these can be simulated by \textsc{DLL-L-UP}.  Hence the ability of
learning large clauses can give a DLL-algorithm a superpolynomial
speedup over one that learns only short clauses.

%%% Local Variables: 
%%% mode: latex
%%% TeX-master: "../main"
%%% End: 

\section{Preliminaries}\label{sec:prelim}

\paragraph{Propositional logic.}
% \label{sec:propLog}
Propositional formulas are formed using Boolean
connectives $\lnot$, $\land$, and $\lor$.  However, this paper works only
with formulas in conjunctive normal form, namely formulas that can be
expressed as a set of clauses.  We write $\overline x$ for the negation
of~$x$, and $\overline{\overline{x}}$ denotes~$x$.
A {\em literal}~$l$ is defined to be
either a variable~$x$ or a negated variable~$\overline x$.  A clause~$C$
is a finite set of literals, and is interpreted as being the disjunction
of its members. The empty clause is denoted~$\Box$. % and is false. 
A {\em unit} clause is a clause containing a single literal. 
A set~$F$ of clauses is interpreted as the conjunction
of its clauses, i.e., a conjunctive normal form formula (CNF).  

An assignment~$\alpha$ is a (partial)
mapping from the set of variables to $\{0,1\}$, where we identify $1$ with {\em True}
and $0$ with {\em False}.  The assignment~$\alpha$
is implicitly extended to assign values to literals
by letting $\alpha(\overline x) = 1-\alpha(x)$,
and the domain, $\dom(\alpha)$, of~$\alpha$ is the set of literals
assigned values by~$\alpha$.
 The {\em restriction} of
a clause~$C$ under~$\alpha$ is the clause
\begin{equation*}
    \rest{C}{\alpha}  =  \left\{ \begin{array}{llll}
                   1 & \text{if there is a } l \in C \text{ with } \alpha(l)=1\\
                   0 & \text{if } \alpha(l)=0 \text{ for every } l \in C\\
                   \set{l \in C}{l \not\in \dom(\alpha)}& \text{otherwise}
              \end{array} \right.
\end{equation*}
The \emph{restriction}
of a set $F$ of clauses under~$\alpha$ is
\begin{equation*}
    \rest{F}{\alpha}  =  \left\{ \begin{array}{llll}
         0 & \text{if there is a } C \in F \text{ with } \rest{C}{\alpha}=0\\
         1 & \text{if } \rest{C}{\alpha}=1 \text{ for every } C \in F\\
         \set{\rest{C}{\alpha}}{C \in F} \setminus \{1\} & \text{otherwise}
  \end{array} \right.
\end{equation*}
If $\rest F \alpha = 1$, then we say $\alpha$ \emph{satisfies}~$F$.

An assignment is called {\em total} if it assigns values to all variables.
We call two CNFs $F$ and~$F^\prime$ \emph{equivalent} and write $F\equiv F^\prime$
to indicate that $F$ and~$F^\prime$ are satisfied by exactly the same 
total assignments.
Note, however, that $F\equiv F^\prime$ does not always imply that
they are satisfied by the same
partial assignments.

If $\epsilon \in \{0,1\}$ and $x$~is a variable, we define $x^\epsilon$ by 
letting $x^0$ be~$x$ and $x^1$ be~$\overline{x}$.

\paragraph{Resolution.}
% \label{sec:resolution}
Suppose that $C_0$ and~$C_1$ are clauses and $x$ is a variable with
$x \in C_0$ and $\overline x \in C_1$.  Then the {\em resolution rule}
can be used to derive the clause
$C = (C_0\setminus\penalty10000\{x\})\cup (C_1\setminus \{\overline x\})$.
In this case we write $C_0,C_1 \vdash_x C$ or just
$C_0,C_1 \vdash C$.

A \emph{resolution proof} of a clause~$C$ from a CNF $F$ consists of
repeated applications of the resolution rule to derive the clause~$C$
from the clauses of~$F$.
If $C = \Box$, then $F$~is unsatisfiable and
the proof is called a {\em resolution refutation}.

We represent resolution proofs either as graphs or as trees.
A {\em resolution dag} (RD) is a dag $G=(V,E)$ with labeled
edges and vertices satisfying the
following properties.
Each node is labeled with a
clause and a variable, and,
in addition, each edge is labeled with a literal.
There must be
a single node of out-degree zero, labeled with 
the conclusion clause.
Further, all nodes with in-degree zero 
are labeled with clauses from the initial 
set~$F$.  All other nodes must have in-degree two and are labeled with a
variable~$x$ and a clause $C$ such that $C_0,C_1\vdash_x C$ where
$C_0$ and~$C_1$ are the labels on the the two immediate
predecessor nodes and $x\in C_0$ and $\overline x\in C_1$.
The edge from $C_0$ to~$C$ is labeled~$\overline x$,
and the edge from $C_1$ to~$C$ is labeled~$x$.  (The convention 
that
that $x\in C_0$ and $\overline x$ is on the edge from~$C_0$
might seem strange,
but it
allows a more natural formulation of
Theorem~\ref{the:regWprops} below.)

A resolution dag~$G$ is \emph{$x$-regular} iff every path in~$G$
contains at most one node that is labeled with the variable~$x$.
$G$~is \emph{regular} (or a
regRD) if $G$ is $x$-regular for every~$x$.

We define the {\em size} of a resolution dag~$G=(V,E)$ to be the 
number $|V|$ of vertices in the dag. 
$\var(G)$ is the set of variables used
as resolution variables in~$G$.  Note that if $G$ is a resolution proof rather
than a refutation, then $\var(G)$ may not include all the variables
that appear in clause labels of~$G$.

A {\em resolution tree} (RT) is a resolution dag which is tree-like,
i.e., a dag in which every vertex other then the conclusion clause has 
out-degree one.
A regular resolution tree is called a regRT for short.

The notion of (p-)simulation is 
an important tool for comparing the strength of proof systems. 
If $\mathcal Q$ and~$\mathcal R$ are refutation systems,
we say that $\mathcal Q$ {\em simulates}~$\mathcal R$
provided there is a polynomial~$p(n)$
such that, for every
~$\mathcal R$-refutation of a CNF~$F$ of size~$n$ there is
a $\mathcal Q$-refutation of~$F$ of size~$\le p(n)$.
If the $\mathcal Q$-refutation can be found by a polynomial time procedure,
then this called a {\em p-simulation}.  Two systems that simulate 
(resp, p-simulate)
each other are called {\em equivalent} (resp, {\em p-equivalent}).
Some basic prior results for simulations of resolution systems include:

\begin{thm}
\hspace*{1ex}
% \label{lem:RT->regRT}
\begin{enumerate}[\em(a)] 
\setlength{\itemsep}{0pt}
\item {\rm \cite{Tseitin68}}
Regular tree resolution (regRT) p-simulates tree resolution (RT).
\item {\rm \cite{Goerdt1993,Alekhnovich2002}}
Regular resolution (regRD) does not simulate resolution (RD).
\item {\rm \cite{BEGJ00}} 
Tree resolution (RT) does not simulate regular resolution (regRD).
\end{enumerate}
\end{thm}

\paragraph{Weakening and w-resolution.}
The {\em weakening} rule
allows the derivation of any clause $C^\prime \supseteq C$
from a clause~$C$.  However, instead of using the weakening
rule, we introduce a {\em w-resolution} rule that essentially incorporates 
weakening into the resolution rule.  Given two clauses $C_0$ and~$C_1$, 
and a variable~$x$, the {\em w-resolution rule} allows one to 
infer
$C = (C_0\setminus\{x\})\cup (C_1\setminus \{\overline x\})$. We
denote this condition $C_0, C_1 \vdash^w_x C$.  Note
that $x\in C_0$ and $\overline x\in C_1$ are not required for the
w-resolution inference.

We use
the notations WRD, regWRD, WRT, and regWRT for the proof systems that
correspond to RD, regRD, RT, and regRT (respectively) but with the resolution
rule replaced with the w-resolution rule.
That is, given a node labeled with $C$, an edge from $C_0$ to $C$ labeled
with $\bar x$ and an edge from $C_1$ to $C$ labeled with $x$, we have
$C = (C_0\setminus\{x\})\cup (C_1\setminus \{\overline x\})$.
% Note that, in spite of its name,
% the weak resolution rule is actually less restrictive than the
% resolution rule.

Similarly, we use the notations RDW and RTW for the proof systems that
correspond to RD and RT, but with the general weakening rule added. In
an application of the weakening rule, the edge connecting a clause
$C^\prime \supseteq C$ with its single predecessor $C$ does not bear
any label.

The resolution and weakening rules
can certainly p-simulate the w-resolution rule,
since a use of the w-resolution rule can be replaced by weakening
inferences that derive $C_0\cup\{x\}$ from $C_0$ and
$C_1\cup\{\overline x\}$ from $C_1$, and then a resolution inference
that derives~$C$.  The converse is not true, since w-resolution
cannot completely simulate weakening; this is because w-resolution
cannot introduce completely new variables that do not occur in the
input clauses.  According to the well-known subsumption principle,
weakening cannot increase the strength of resolution though, and the
same reasoning implies the same about w-resolution; namely, we
have:

\begin{prop}\label{pro:subsumeweak}
  Let $R$ be a WRD proof of~$C$ from~$F$ of size~$n$.  Then there is an
  RD proof~$S$ of~$C^\prime$ from~$F$ of size $\le n$ for
  some $C^\prime\subseteq C$.  Furthermore, if $R$ is regular, so
  is~$S$, and if $R$ is a tree, so is~$S$.
\end{prop}
\proof
The proof of the theorem is straightforward.  Writing $R$ as a
sequence $C_0, C_1, \ldots, C_n = C$, define clauses $C_i^\prime
\subseteq C_i$ by induction on~$i$ so that the new clauses form the
desired proof~$S$.  For $C_i\in F$, let $C^\prime_i =C_i$.  Otherwise
$C_i$~is inferred by w-resolution from $C_j$ and~$C_k$ w.r.t.\ a
variable~$x$.  If $x\in C_j$ and $\overline x \in C_k$, let
$C_i^\prime$ be the resolvent of $C_j^\prime$
and~$C_k^\prime$ as obtained by the usual resolution
rule; if not, then let $C^\prime_i$ be $C^\prime_j$ if
$x\notin C^\prime_j$, or~$C^\prime_k$ if $\overline x \notin
C^\prime_k$.  It is easy to check that each $C_i^\prime \subseteq C_i$
and that, after removing duplicate clauses, the clauses~$C^\prime_j$
form a valid resolution proof~$S$.  If $R$~is regular, then so is~$S$,
and if $R$~is a tree so is~$S$.
\qed

Essentially the same proof shows the same property for the system with the
full weakening rule:
\begin{prop}\label{pro:rtweak}
  Let $R$ be a RDW proof of~$C$ from~$F$ of size~$s$.  Then there is an
  RD proof~$S$ of~$C^\prime$ from~$F$ of size $\le s$ for
  some $C^\prime\subseteq C$.  Furthermore, if $R$ is regular, so is~$S$,
  and if $R$ is a tree, so is~$S$.
\end{prop}

There are several reasons why we prefer to work with w-resolution, rather
than with the weakening rule.  First, we find it to be an elegant
way to combine weakening with resolution.
Second, it works well for using resolution
trees (with input lemmas, see the next section) to simulate DLL search
algorithms.  Third, since weakening and resolution together are stronger than
w-resolution, w-resolution is a more refined restriction on
resolution.  Fourth, for regular resolution, using 
w-resolution instead of general weakening can be a quite restrictive condition,
since any w-resolution inference
$C_0, C_1 \wres_x C$
``uses up'' the variable~$x$, making it unavailable for other
resolution inferences on the same path, even if the variable does not
occur at all in $C_0$ and~$C_1$.  The last two reasons mean that
w-resolution can be rather weak; this strengthens
our results below 
(Theorems \ref{the:regWRTIforLearnables} and~\ref{the:rWRTIsimDLL})
about the existence
of regular proofs that use w-resolution.

The following simple theorem gives some
useful properties for regular w-resolution.
\begin{thm}\label{the:regWprops}
Let $G$ be a regular w-resolution refutation. Let $C$ be a clause in~$G$.
\begin{enumerate}[\em(a)]
\item Suppose that $C$~is
derived from $C_0$ and~$C_1$ with the edge from~$C_0$ (resp.~$C_1$)
to~$C$ labeled with~$\overline x$ (resp.~$x$). Then $\overline x\notin C_0$,
and $x\notin C_1$.
\item
Let $\alpha$ be an assignment such that for every literal~$l$ labeling
an edge on the path from~$C$ to the final clause, $\alpha(l) = True$.
Then $\rest C \alpha = 0$.
\end{enumerate}
\end{thm}
\proof
The proof of part a.~is based on the observation that if $\overline x \in C_0$,
then also $\overline x \in C$.  However, by the regularity of the
resolution refutation, every clause on the path from~$C$ to the final
clause~$\Box$ must contain~$\overline x$.  But clearly $\overline x\notin\Box$.

Part b.~is a well-known fact for regular resolution proofs. It holds
for similar reasons for regular w-resolution proofs: the proof proceeds
by induction on clauses in the proof, starting at the final clause~$\Box$
and moving up towards the leaves.  Part~a.\ makes the induction step trivial.
\qed

\paragraph{Directed acyclic graphs}
We define some basic concepts that will be useful
for analyzing both resolution proofs and
conflict graphs (which are defined below in \pref{sec:dll-up}). 
Let $G=(V,E)$ be a dag.
The set of leaves (nodes in~$V$ of in-degree~0) of~$G$ is denoted $\leafs{G}$.
The {\em depth} of a node~$u$ in~$V$ is defined
to equal the maximum number of edges
on any path from a leaf of~$G$ to the node~$u$.  Hence leaves have depth~$0$.
The subgraph rooted at~$u$ in~$G$ is denoted $\graph u G$; its nodes
are the nodes~$v$ for which there is a path from $v$ to~$u$ in~$G$, and its
edges are the induced edges of~$G$.

%%% Local Variables: 
%%% mode: latex
%%% TeX-master: "../main"
%%% End: 

\section{w-resolution trees with lemmas}\label{sec:wrtl}

This section first gives an alternate characterization of resolution
dags by using \emph{resolution trees with lemmas}.  We then refine the notion 
of lemmas to allow only \emph{input lemmas}.  For non-regular derivations,
resolution trees with lemmas and resolution trees with input lemmas
are both proved below to be p-equivalent to resolution. 
However, for regular proofs,
the notions are apparently different.  (In fact we give an exponential
separation between regular resolution and regular w-resolution trees with
input lemmas.)  Later in the paper we will
give a tight correspondence between resolution trees with input lemmas
and DLL search algorithms.

% Reinhold Letz \cite{Letz2007}.

The intuition for the definition of a resolution tree with lemmas is
to allow any clause proved earlier in the resolution tree to be reused as
a leaf clause.  More formally, assume we are
given a resolution proof tree~$T$, and further assume~$T$ is {\em ordered} in
that each internal node has a left child and a right child.
We define
$ <_T $ to be the post-ordering of~$T$,
namely, the linear ordering of the nodes of~$T$
such that if $u$~is a node in~$T$ and $v$~is in the subtree rooted 
at $u$'s left child, and $w$~is in the subtree rooted at $u$'s right
child, then $v <_T w <_T u$.
For $F$ a set of clauses, a {\em resolution tree with lemmas} (RTL) proof
from~$F$
is an ordered binary tree such that
(1)~each leaf node~$v$ is labeled with either a member
of~$F$ or with a clause that labels some node $u <_T v$, and
(2)~each internal node~$v$ is labeled with a variable~$x$ and a clause~$C$, 
such that~$C$
is inferred by resolution w.r.t.~$x$ from the clauses labeling the two children
of~$v$, and (3)~the unique out-degree zero node is labeled
with the conclusion clause~$D$.  If $D=\Box$, then the RTL proof is
a refutation.

{\em w-resolution trees with lemmas} (WRTL) are defined just like
RTL's, but allowing w-resolution in place of resolution, and
\emph{resolution trees with lemmas and weakening} (RTLW) are defined
in the same way, but allowing the weakening rule in addition to
resolution.

An RTL or WRTL proof is {\em regular} provided 
that no path in the proof tree contains more than one (w-)resolution
using a given variable~$x$.  Note that paths follow the tree edges
only; any maximal path starts at a leaf node (possibly
a lemma) and ends at the conclusion.

It is not hard to see that resolution trees with lemmas (RTL) and resolution dags (RD)
p-simulate each other.  Namely, an RD can be converted into an RTL by doing
a depth-first, leftmost traversal of the RD.  In addition, it is clear
that regular RTL's p-simulate regular RD's.  The converse
is open,
and it is false for regular WRTL, as we prove in \pref{sec:regwrti-dll}:
intuitively, the problem is that
when one converts an RTL proof into an RD, new path connections are
created when leaf clauses are replaced with edges back to the node
where the lemma was derived.

We next define resolution trees with input lemma (RTI) proofs.  These
are a restricted version of resolution trees with lemmas, where the lemmas
are required to have been derived earlier in the proof by \emph{input proofs}.
Input proofs have also been called \emph{trivial proofs} by
Beame et al.~\cite{BeameKautzSabharwal2004}, and they are useful for 
characterizing the clause learning permissible for DLL algorithms.
\begin{defi}
An {\em input resolution tree} is a resolution tree such 
that every internal node
has at least one child that is a leaf. 
Let $v$~be a node
in a tree~$T$ and let $T_v$ be the subtree of~$T$ with root~$v$.
The node~$v$ is called
an {\em input-derived node} if $T_v$~is an input resolution tree.
\end{defi}

Often the node~$v$ and its label~$C$ are identified.  In this case,
$C$~is called an {\em input-derived clause}.  In RTI proofs,
input-derived clauses may be reused as lemmas.  Thus, in an RTI proof,
an input-derived clause is derived by an input proof whose leaves 
either are initial clauses or are clauses that were already input-derived.

\begin{defi}
A {\em resolution tree with input lemmas} (RTI) proof~$T$
is an RTL proof with the
extra condition that every lemma in~$T$ must appear earlier in~$T$ as
an input-derived clause.  That is to say, every leaf node~$u$ in~$T$ is
labeled either with an initial clause from~$F$ or with a clause that labels
some input-derived node $v <_T u$.
\end{defi}
The notions of w-resolution trees with input lemmas (WRTI), regular
resolution trees with input lemmas (regRTI), and regular w-resolution
trees with input lemmas (regWRTI) are defined similarly.%
\footnote{A small, but
important point is that w-resolution inferences are not allowed in
input proofs, even for input proofs that are part of WRTI proofs.  
We have chosen the definition of input proofs so as
to make the results in \pref{sec:regwrti-dll} hold that show the equivalence
between regWRTI proofs and DLL-L-UP search algorithms.
Although similar results could be
obtained if the definition of input proof were changed to allow 
w-resolution inferences, it would require also using a modified, and
less natural, version of clause learning.}

It is clear that the resolution dags (RD) and resolution trees with lemmas (RTL)
p-simulate resolution trees with input lemmas (RTI).  Somewhat surprisingly,
the next theorem shows that the converse p-simulation holds as well.

\begin{thm}\label{the:RD->RTI}
  Let $G$ be a resolution dag of
  size~$s$ for the clause~$C$ from the set~$F$ of clauses. 
Let $d$ be the depth of~$C$ in~$G$.
Then there is an RTI proof~$T$ for~$C$ from~$F$ of size $< 2sd$.
If $G$ is regular then $T$ is also regular.
\end{thm}

\proof
The dag proof~$G$ can be unfolded into a proof tree~$T^\prime$, possibly
exponentially bigger.  The proof idea is to prune clauses away 
from~$T^\prime$ leaving a RTI proof~$T$ of the desired size.

Without loss of generality, no clause appears more than once in~$G$; hence,
for a given clause~$C$ in the tree~$T^\prime$, every occurrence of~$C$
in~$T^\prime$ is derived by the same subproof~$T^\prime_C$. 
Let $d_C$ be the depth of~$C$
in the proof, i.e., the height of the tree~$T^\prime_C$.  Clauses
at leaves have depth~$0$.
We give the proof tree~$T^\prime$ an arbitrary left-to-right order, so that it
makes sense to talk about the $i$-th occurrence of a clause~$C$ in~$T^\prime$.

We define the
$j$-th occurrence of a clause~$C$ in~$T^\prime$ to be \emph{leafable},
provided $j > d_C$.  The intuition is that the leafable clauses
will have been proved as a input clause earlier in~$T$, and thus
any leafable clause may be used as a lemma in~$T$.  

To form~$T$ from~$T^\prime$, remove from~$T^\prime$ any clause~$D$ if it has a
successor that is leafable, so that every leafable occurrence of a clause 
either does not appear in~$T$ or appears in~$T$
as a leaf.
To prove that $T$~is a valid RTI proof, it suffices
to prove, by induction on~$i$, that if $C$~has depth $d_C=i>0$,
then the
$i$-th occurrence of~$C$ is input-derived in~$T$.
Note that the two children $C_0$ and~$C_1$ of~$C$
must have depth $<d_C$.
Since every occurrence of~$C$ is derived from the same two clauses, these
occurrences of $C_0$ and~$C_1$ must be at least their
$i$-th occurrences.  Therefore, by the induction hypothesis, the
children $C_0$ and~$C_1$ are leafable and appear in~$T$ as leaves.
Thus, since it is derived by a single
inference from two leaves, the $i$-th occurrence of~$C$ is input-derived.

It follows that $T$ is a valid RTI proof.  If the proof~$G$ was regular, 
clearly $T$~is regular too.

To prove the size bound for~$T$,
note that $G$ has at most
$s-1$ internal nodes.  Each one occurs at most $d$ times as an internal
node in~$T$, so $T$ has at most $d(s-1)$ internal nodes.  Thus, $T$~has
at most $2d\cdot (s-1) +1 < 2sd$ nodes in all.
\qed

The following two theorems summarize the relationships between our
various proof systems.  We write ${\mathcal R}\equiv{\mathcal Q}$ to denote
that $\mathcal R$ and~$\mathcal Q$ are p-equivalent, and ${\mathcal Q}\le {\mathcal R}$
to denote that $\mathcal R$ p-simulates $\mathcal Q$.  The notation
${\mathcal Q} < {\mathcal R}$ means that $\mathcal R$ p-simulates~$\mathcal Q$ but
$\mathcal Q$ does not simulate~$\mathcal R$.
\begin{thm}\label{the:all_equiv}
  $\text{RD} \equiv \text{WRD} \equiv \text{RTI} \equiv \text{WRTI}
  \equiv \text{RTL} \equiv \text{WRTL}$
\end{thm}
\proof
The p-equivalences
  $\text{RD} \equiv \text{WRD}$ and $\text{RTI} \equiv \text{WRTI}$ and
  $\text{RTL} \equiv \text{WRTL}$ are shown by (the proof of)
\pref{pro:subsumeweak}.  The simulations 
$\text{RTI} \le \text{RTL} \equiv \text{RD}$ are straightforward.
Finally, $\text{RD} \le \text{RTI}$ is shown by Theorem~\ref{the:RD->RTI}.
\qed

For regular resolution, we have the following theorem.

\begin{thm}\label{the:hierarchy}
  $\text{regRD} \equiv \text{regWRD} \leq \text{regRTI}
  \leq \text{regRTL} \leq \text{regWRTL} \leq \text{RD}$ and
  $\text{regRTI} \leq \text{regWRTI} \leq \text{regWRTL}$.
\end{thm}

\proof
  $\text{regRD} \equiv \text{regWRD}$ and $\text{regWRTL} \leq \text{RD}$
follow from the definitions and the proof of \pref{pro:subsumeweak}.
The p-simulations  $\text{regRTI} \leq \text{regRTL}
  \leq \text{regWRTL}$ and $\text{regRTI} \leq
  \text{regWRTI} \leq \text{regWRTL}$ follow from the definitions.
The p-simulation $\text{regRD} \leq \text{regRTI}$ is shown by
Theorem~\ref{the:RD->RTI}.
\qed

Below, we prove, as \pref{the:regRDnosimregWRTI}, that
$\text{regRD} < \text{regWRTI}$.
This is the only separation in the hierarchy that is known.
In particular, it is open whether
$\text{regRD} < \text{regRTI}$,
$\text{regRTI} < \text{regRTL}$, $ \text{regRTL} < \text{regWRTL}$, $
\text{regWRTL} < \text{RD}$ or $\text{regWRTI} < \text{regWRTL}$ hold.
It is also open whether regWRTI and regRTL are comparable.

%%% Local Variables: 
%%% mode: latex
%%% TeX-master: "../main"
%%% End: 

\section{DLL algorithms with clause learning}
\label{sec:dll-up}

\subsection{The basic DLL algorithm}
\label{sec:basic_dll}

The DLL proof search algorithm is named after the
authors Davis, Logeman and Loveland of 
the paper where it was introduced~\cite{DavisLogemann1962}.  Since they built
on the work of Davis and Putnam~\cite{DavisPutnam1960}, the algorithm
is sometimes called the DPLL algorithm.
There are several variations on the DLL algorithm, but the basic
algorithm is shown in \pref{fig:dll}.  The input is a set~$F$ of
clauses, and a partial assignment~$\alpha$.  The assignment~$\alpha$
is a set of ordered pairs $(x,\epsilon)$, where $\epsilon\in\{0,1\}$,
indicating that $\alpha(x)=\epsilon$.
The DLL algorithm is implemented
as a recursive procedure and returns
either
\texttt{UNSAT} if $F$ is unsatisfiable or otherwise a satisfying assignment
for~$F$.

\begin{figure}[htbp]
\begin{center}
\begin{minipage}{1.0\linewidth}
 \tt \small
  \begin{tabbing}
    123\=123455\=12345\=12345\=12345\=12345\=12345\=12345 \kill
    \>{\sc DLL}($F,\alpha$)\\
    \>1\>if $\rest{F}{\alpha} = 0$ then\\
    \>2\>\>return UNSAT\\
    \>3\>if $\rest{F}{\alpha} = 1$ then\\
    \>4\>\>return $\alpha$ \\
    \>5\>choose $x \in \var(\rest{F}{\alpha})$ and $\epsilon\in\{0,1\}$ \\
    \>6\>$\beta \leftarrow${\sc DLL}($F,\alpha \cup \{(x,\epsilon)\}$)\\
    \>7\>if $\beta \neq$ UNSAT then\\
    \>8\>\>return $\beta$\\
    \>9\>else\\
    \>10\>\>return {\sc DLL}($F,\alpha \cup \{(x,1-\epsilon)\}$)
  \end{tabbing}
\end{minipage}
\caption{The basic DLL algorithm.}
\label{fig:dll}
\end{center}
\end{figure}

Note that the DLL algorithm is not fully specified, since line~5 does not
specify how
to choose the branching variable~$x$
and its value~$\epsilon$.
Rather one can think of the algorithm either as being nondeterministic or as
being an algorithm schema.  
We prefer to think of the algorithm as an algorithm schema, so that it
incorporates a variety of possible algorithms.  Indeed, there has
been extensive research into how to choose the branching variable
and its value \cite{Freeman1995,Nadel2002}.

There is a well-known close connection between regular
resolution and DLL algorithms.
In particular, a run of DLL can be viewed as a regular resolution
tree, and vice-versa.  This can be formalized by the following two propositions.

\begin{prop} \label{pro:DLL_RT2} Let $F$ be an unsatisfiable set of
clauses and $\alpha$~an
  assignment.  If there is an execution of \hbox{\rm{\sc DLL}($F,\alpha$)}
  that
  returns \texttt{UNSAT} and performs $s$ recursive calls, then there
  exists a clause $C$ with $\rest{C}{\alpha} = 0$ such that $C$~has a
  regular resolution tree~$T$ from~$F$ with $|T| \leq s+1$ and
  $\var(T) \cap \dom(\alpha) = \varnothing$.
\end{prop}

The converse simulation of \pref{pro:DLL_RT2} holds, too, that is, a 
regular resolution tree can be transformed directly in a run of
\textsc{DLL}. 

\begin{prop} \label{pro:DLL_RT1}
Let $F$ be an unsatisfiable 
set of clauses.
 Suppose that $C$~has
  a regular resolution proof tree~$T$ of size~$s$ from~$F$.
  Let $\alpha$ be an assignment with $\rest{C}{\alpha} = 0$ and
  $\var(T) \cap \dom(\alpha) = \varnothing$.  Then there is an
  execution of \hbox{\rm{\sc DLL}($F,\alpha$)}, that returns \texttt{UNSAT}
  after at most $s-1$ recursive calls.
\end{prop}

The two propositions are based on the following correspondence between
resolution trees and a DLL search tree: first, a leaf clause in
a resolution tree corresponds to a clause falsified by~$\alpha$ (so that
$\rest F \alpha = 0$), and second, a resolution inference with respect to
a variable~$x$ corresponds to the use of $x$~as a
branching variable in the DLL algorithm.
Together the two propositions give the following
well-known exact correspondence between regular
resolution trees and DLL search.

\begin{thm}\label{the:DLL_RT}
  If $F$ is unsatisfiable, then there is an execution of 
\hbox{\rm {\sc DLL}($F,\varnothing$)} that executes with $< s$ recursive calls
  if and only if there exists a regular refutation tree for~$F$ of
size $\le s$.
\end{thm}

\subsection{Learning by unit propagation}

Two of the most successful enhancements of DLL that are used by most
modern SAT solvers are unit propagation and clause learning.  
\emph{Unit clause propagation} (also called Boolean
constraint propagation) was already part of the original DLL algorithm
and is based on the following observation: If
$\alpha$ is a partial assignment for a set of clauses~$F$ and if there
is a clause $C\in F$ with $\rest C \alpha = \{ l \}$ a unit clause,
then any $\beta\supset \alpha$ that satisfies~$F$ must assign $l$ the
value {\em True}.

There are a couple of methods that the DLL algorithm can 
use to implement unit propagation.
One method is to just use unit propagation
to guide the choice of a branching variable by modifying line~5 so that,
if there is a unit clause in~$\rest F \alpha$, then $x$ and~$\epsilon$
are chosen to make the literal true.  More commonly though, DLL algorithms
incorporate unit propagation as a separate phase during which the
assignment~$\alpha$ is iteratively extended to make any unit clause
true until there are no unit clauses remaining.  As the unit propagation
is performed, the DLL algorithm keeps track of which variables were
set by unit propagation and which clause was used as the basis for
the unit propagation.  This information is then useful for clause
learning.

\emph{Clause learning} in DLL algorithms was first introduced
by Silva and Sakallah~\cite{SilvaSakallah1996} and  
means that
new clauses are effectively added to~$F$.
A learned clause~$D$ must be implied by~$F$, so that adding $D$ to~$F$
does not change the space of satisfying assignments.
In theory, there are many potential methods for clause learning; however,
in practice, the only useful method for learning clauses is based on
unit propagation as in the original proposal \cite{SilvaSakallah1996}.
In fact, all deterministic state of the art
SAT solvers for structured (non-random) instances of SAT are
based on clause learning via unit propagation.  This includes
solvers such as Chaff~\cite{MoskewiczMalik2001}, 
Zchaff~\cite{MahajanFu2004} and MiniSAT~\cite{EenBiere2005}.

These DLL algorithms apply clause learning when the set~$F$
is falsified by the current assignment~$\alpha$.  Intuitively,
they analyze the {\em reason} some clause~$C$ in~$F$ is falsified
and use this reason to infer a
clause~$D$ from~$F$ to be learned.  There are two ways
in which a DLL algorithm assigns values to variables, namely, by unit
propagation and by setting a branching variable.  However, if unit propagation
is fully carried out, then the first time a clause is falsified is during
unit propagation.  In particular, this happens when there are two unit
clauses $\rest{C_1}\alpha = \{x \}$ and $\rest{C_2}\alpha = \{ \overline x \}$
requiring a variable~$x$ to be set both {\em True} and {\em False}.  This
is called a {\em conflict}.

The reason for a conflict is analyzed by building a 
conflict graph.   Generally, this is done by maintaining an \emph{unit
propagation graph} that tracks, for each variable which
has been assigned a value, the reason that implies the setting of the variable.
The two possible reasons are that either (a)~the variable was set by
unit propagation when a
particular clause~$C$ became a unit clause, in which case $C$~is the reason,
or (b)~the variable was set arbitrarily as a branching variable.  The
unit propagation graph~$G$ has literals as its nodes.  The leaves of~$G$
are literals that were set true as branching variables,
and the internal nodes are variables that were set true
by unit propagation.  If a literal~$l$ is an internal node in~$G$,
then it was set
true by unit propagation applied to some clause~$C$.  In this case, for
each literal~${l^\prime}\not= l$ in~$C$, $\overline{l^\prime}$~is
a node in~$G$ and there
is an edge from $\overline{ l^\prime }$ to~$l$.  If the
unit propagation graph contains a conflict it is called a \emph{conflict graph}.
More formally, a conflict graph is defined as follows.
\begin{defi}
A {\em conflict graph}~$G$ for a set~$F$ of clauses
under the assignment~$\alpha$
is a 
dag $G=(V\cup\{\Box\},E)$
where $V$ is a set of literals and where the following hold:
\begin{enumerate}[(a)]
\setlength{\itemsep}{1pt}
\item For each $l\in V$, either (i)~$l$ has in-degree~0 and
$\alpha(l)=1$,
or (ii)~there is
a clause~$C\in F$ such that 
$C = \{l\} \cup \{ l^\prime : (\overline {l^\prime},l)\in E\}$.
For a fixed conflict graph~$G$, we denote this clause as~$C_l$.
\item There is a unique variable~$x$ such that 
$V\supseteq\{x,\overline x\}$.
\item The node~$\Box$ has only the two incoming edges 
$(x,\Box)$ and $(\overline x,\Box)$.
\item The node $\Box$ is the only node with outdegree zero.
\end{enumerate}
\end{defi}

Let $\leafs{G}$ denote the nodes in~$G$ of in-degree zero.  Then, letting
$\alpha_G = \{ (x,\epsilon) : x^\epsilon \in \leafs G \}$, the conflict
graph~$G$ shows that every vertex~$l$ must be made true 
by any satisfying assignment for~$F$ that extends~$\alpha$.  Since 
for some~$x$, both $x$ and $\overline x$ are nodes of~$G$, this implies
$\alpha$ cannot be extended to a satisfying assignment for~$F$.  
Therefore, the clause $D = \{ \overline l : l\in\leafs G \}$ is implied
by~$F$, and $D$~can be taken as a learned clause.  We call this clause~$D$
the {\em conflict clause} of~$G$ and denote it $\Cc G$.

There is a second type of clause that can be learned from the conflict
graph~$G$ in
addition to the conflict clause~$\Cc G$.  Namely, let $l\not = \Box$
be any non-leaf node
in~$G$.  Further, let $\leafs { \graph l G }$ be the set of leaves~$l^\prime$
of~$G$
such that there is a path from~$l^\prime$ to~$l$.  Then, the clauses in~$F$
imply that if all the leaves~$l^\prime \in \leafs{\graph l G}$ are assigned
true, then $l$~is assigned true. Thus, the clause 
$D = \{ l \} \cup
\{ \overline {l^\prime} : l^\prime \in \leafs{\graph l G} \}$ 
is
implied by~$F$ and can be taken as a learned clause.  This clause~$D$
is called the {\em induced clause} of $G_l$ and is denoted $\ic l G$.
In the degenerate case where $\graph l G$ consists of only the single
literal~$l$, this would make $\ic l G$ equal to $\{ l, \overline l \}$; rather
than permit this as a clause, we instead say that the induced clause does
not exist.

In practice, both conflict clauses $\Cc G$ and induced clauses~$\ic l G$
are used by SAT solvers.  It appears that most SAT solvers learn the
\emph{first-UIP} clauses~\cite{SilvaSakallah1996}, which equal $\Cc{G}$ 
and $\ic l {G^\prime}$ for appropriately formulated~$G$ and~$G^\prime$.
Other conflict clauses that can be learned include
\emph{all-UIP} clauses~\cite{ZhangMadigan2001}, 
\emph{rel-sat} clauses~\cite{BayardoSchrag:CSPlookback}, 
\emph{decision} clauses~\cite{ZhangMadigan2001},
and
\emph{first cut} clauses~\cite{BeameKautzSabharwal2004}.
All of these are conflict clauses $\Cc{G}$ for appropriate~$G$.
Less commonly, multiple clauses are learned, including clauses
based on the cuts
advocated by the mentioned works \cite{SilvaSakallah1996,ZhangMadigan2001}, which 
are a type of induced clauses.

In order to prove the correspondence in \pref{sec:regwrti-dll} between
DLL with clause learning and regWRTI proofs, we must put some restrictions
on the kinds of clauses that can be (simultaneously)
learned.  In essence, the point is that for DLL with clause learning to
simulate regWRTI proofs it is necessary to learn multiple clauses at
once in order to learn all the clauses in a regular input subproof.
But on the other hand, for regWRTI to simulate DLL with clause learning,
regWRTI must be able to include regular input proofs that derive
all the learned clauses so as to have them available for subsequent use as
input lemmas.  Thus, we define a notion of ``compatible clauses'' which 
is a set of clauses that can be simultaneously learned.  For this, we
define the notion of a series-parallel decomposition of a conflict graph~$G$.

\begin{defi}
A graph~$H=(W,E^\prime)$ is a {\em subconflict graph} of
the conflict graph~$G=(V,E)$ provided that
$H$~is a conflict graph with $W\subseteq V$ and $E^\prime\subseteq E$,
and that each non-leaf vertex of~$H$ (that is,
each vertex in $W\setminus \leafs H$)
has the same in-degree in~$H$ as in~$G$.

$H$~is a {\em proper} subconflict graph of~$G$ provided 
there is no path in~$G$ from any non-leaf vertex of~$H$
to a vertex in~$\leafs H$.
\end{defi}
Note that if $l$~is a non-leaf vertex in the subconflict graph~$H$ of~$G$,
then the clause $C_l$ is the same whether it is defined with respect to~$H$ or 
with respect to~$G$.
\begin{defi}
Let $G$~be a conflict graph.  A 
{\em decomposition} of~$G$ is a sequence
$H_0\subset H_1\subset \cdots\subset H_k$, $k\ge 1$, of distinct
proper
subconflict graphs of~$G$ such that $H_k=G$ and 
$H_0$~is the dag on the three nodes~$\Box$ and its
two predecessors $x$ and~$\overline x$.
\end{defi}
A decomposition of~$G$ will be used to describe sets of clauses
that can be simultaneously learned.  For this, we put a structure
on the decomposition that describes the exact types of clauses
that can be learned:
\begin{defi}
A {\em series-parallel decomposition}~$\mathcal H$ of~$G$ consists of a
decomposition $H_0,\ldots,H_k$ plus, for each $0\le i<k$, 
a sequence
$H_i=H_{i,0}\subset H_{i,1}\subset \cdots \subset H_{i,m_i}=H_{i+1}$
of proper subconflict graphs of~$G$.
Note that the sequence
\[
H_0=H_{0,0}, H_{0,1}, H_{0,2},\ldots,
 H_{0,m_0}=H_1=H_{1,0}, H_{1,1}, \ldots,
 H_{k-1,m_{k-1}} = H_k
\]
is itself a decomposition of~$G$.  However, we prefer to view
it as a two-level decomposition.
A {\em series} decomposition is a series-parallel decomposition with
trivial parallel part, i.e., with $k=1$.
A {\em parallel} decomposition is series-parallel decomposition
in which $m_i=1$ for all~$i$.
Note that
we always have $H_i\not= H_{i+1}$
and $H_{i,j}\not=H_{i,j+1}$.
\end{defi}
Figure~\ref{seriesparallelFig} illustrates a series-parallel decomposition.
\begin{defi}
For $\mathcal H$ a series-parallel decomposition, the set of {\em learnable
clauses}, $\Cc {\mathcal H}$, for~$\mathcal H$ consists of the following
induced clauses and conflict clauses:
\begin{enumerate}[$\bullet$]
\item For each $1\le j \le m_0$, the conflict clause
$\Cc {H_{0,j}}$, and
\item For each $0<i<k$ and $0<j\le{m_i}$
and each $l \in \leafs{H_i} \setminus \leafs{H_{i,j}}$, the induced
clause $\ic l {H_{i,j}}$.
\end{enumerate}
\end{defi}

\begin{figure}[t]
\psset{unit=0.04cm}
\begin{center}
\begin{pspicture}(-50,0)(175,200)
\cnodeput(0,0){box}{\makebox(0, 6.6){$\Box$}}
\cnodeput(20,20){abar}{\makebox(0, 6.6){$\overline a$}}
\cnodeput(-20,20){a}{\makebox(0, 6.6){$a$}}
\ncline{a}{box}
\ncline{abar}{box}
\cnodeput(20,40){c}{\makebox(0, 6.6){$c$}}
\cnodeput(-20,50){b}{\makebox(0, 6.6){$b$}}
\cnodeput(20,60){d}{\makebox(0, 6.6){$d$}}
\cnodeput(0,80){e}{\makebox(0, 6.6){$e$}}
\ncline{c}{abar}
\ncline{b}{a}
\ncline{b}{abar}
\ncline{d}{c}
\ncline{e}{b}
\ncline{e}{d}
\cnodeput(-20,105){f}{\makebox(0, 6.6){$f$}}
\cnodeput(20,105){g}{\makebox(0, 6.6){$g$}}
\cnodeput(-25,130){h}{\makebox(0, 6.6){$h$}}
\cnodeput(15,130){i}{\makebox(0, 6.6){$i$}}
\ncline{f}{e}
\ncline{g}{e}
\ncline{h}{f}
\ncline{i}{f}
\ncline{i}{g}
\cnodeput(-20,160){j}{\makebox(0, 6.6){$j$}}
\cnodeput(20,160){k}{\makebox(0, 6.6){$k$}}
\cnodeput(-5,180){ell}{\makebox(0, 6.6){$\ell$}}
\cnodeput(25,180){m}{\makebox(0, 6.6){$m$}}
\ncline{j}{h}
\ncline{j}{i}
\ncline{k}{i}
\ncline{ell}{j}
\ncline{ell}{k}
\ncline{m}{k}
\psline[linewidth=1.5pt](-30,30)(30,30)
\psline[linewidth=1.5pt](-30,92)(30,92)
\psline[linewidth=1.5pt](-30,147)(30,147)
\psline[linewidth=1.5pt](-30,195)(30,195)
\pscurve[linewidth=1.5pt,linestyle=dotted](-30,62)(-25,62)(20,50)(30,50)
\pscurve[linewidth=1.5pt,linestyle=dotted](-30,67)(-25,67)(25,72)(30,72)
\psline[linewidth=1.5pt,linestyle=dotted](-30,115)(30,115)
\pscurve[linewidth=1.5pt,linestyle=dotted]%
(-30,120)(-25,120)(10,138)(20,140)(30,140)
\rput[l](33,30){$H_0 = H_{0,0}$}
\rput[l](33,50){$H_{0,1}$}
\rput[l](33,72){$H_{0,2}$}
\rput[l](33,92){$H_1=H_{0,3}=H_{1,0}$}
\rput[l](33,115){$H_{1,1}$}
\rput[l](33,137){$H_{1,2}$}
\rput[l](33,147){$H_2=H_{1,3}=H_{2,0}$}
\rput[l](33,194){$H_3=H_{2,1}$}
\rput[c](150,200){\underline{Learnable clauses}}
\rput[c](150,188){$\{ \overline\ell, h \}$ }
\rput[c](150,175){$\{ \overline\ell, \overline m, i \}$ }
\rput[c](150,150){$\{ \overline h, \overline i, e \}$ }
\rput[c](150,138){$\{ \overline f, \overline i, e \}$ }
\rput[c](150,115){$\{ \overline f, \overline g, e \}$ }
\rput[c](150,92){$\{ \overline e \}$ }
\rput[c](150,72){$\{ \overline b, \overline d \}$ }
\rput[c](150,50){$\{ \overline b, \overline c \}$ }
\end{pspicture}
\end{center}
\caption{A series-parallel decomposition.  Solid lines define
the sets~$H_i$ of the parallel part of the decomposition, and dotted lines
define the sets $H_{i,j}$ in the series part.  Each line (solid or dotted)
defines the set of nodes that lie below the line.
The learnable clauses associated with each
set are shown in the right column.
}
\label{seriesparallelFig}
\end{figure}

It should be noted that the definition of the parallel decomposition
incorporates the notion of ``cut'' used 
by Silva and Sakallah~\cite{SilvaSakallah1996}.  
The DLL algorithm shown in \pref{fig:dll-l-up} chooses
a single series-parallel decomposition~$\mathcal H$
and learns some subset of the learnable clauses in~$\Cc {\mathcal H}$.
It is clear
that this generalizes all of the clause learning algorithms
mentioned above.

The algorithm schema \textsc{DLL-L-UP} that is given in
\pref{fig:dll-l-up} is a modification of the schema \textsc{DLL}.  In
addition to returning a satisfying assignment
or \texttt{UNSAT}, it returns a modified formula that might include
learned clauses. If $F$ is a set of clauses 
and $\alpha$~is an assignment then \textsc{DLL-L-UP}($F,\,\alpha$)
returns $(F',\alpha')$ such that $F^\prime \supseteq F$ and
$F^\prime$ is equivalent to~$F$ and such that
$\alpha^\prime$~either
is \texttt{UNSAT} or is a satisfying assignment for~$F$.\footnote{
Our definition of \textsc{DLL-L-UP} is slightly different from the version of
the algorithm as originally defined in Hoffmann's thesis \cite{Hoffmann2007}.  The first
main difference is that we use series-parallel decompositions
rather the compatible set of subconflict graphs of Hoffmann~\cite{Hoffmann2007}. 
The second 
difference is that our algorithm does not build the implication 
graph incrementally by the use of explicit unit propagation;
instead, it builds the implication graph once
a conflict has been found.}

\begin{figure}[htbp]
\begin{center}
\begin{minipage}{1.0\linewidth}
 \tt \small
  \begin{tabbing}
  123\=123455\=12345\=12345\=12345\=12345\=12345\=12345\=12345\=12345\=12345\=12345 \kill
  \>{\sc DLL-L-UP}($F,\alpha$)\\
  \>1\>if $\rest{F}{\alpha} = 1$ then return ($F,\alpha$) \\
  \>2\>if there is a conflict graph for~$F$ under~$\alpha$ then \\
  \>3\>\>choose a conflict graph~$G$ for~$F$ under~$\alpha$ \\
  \>4\>\>\>and a series-parallel decomposition~$\mathcal H$ of~$G$  \\
  \>5\>\>choose a subset $S$ of $\Cc{{\mathcal H}}$ ~~ -- the learned clauses \\
  \>6\>\>return ($F\cup S$, UNSAT) \\
  \>7\>choose $x \in \var(\rest{F}{\alpha})$ and $\epsilon \in \{0,1\}$\\
  \>8\>($G,\beta$)$\leftarrow${\sc DLL-L-UP}($F,\alpha \cup \{(x,\epsilon)\}$)\\
  \>9\>if $\beta \neq $ UNSAT then \\
  \>10\>\>return ($G,\beta$)\\
  \>11\>return {\sc DLL-L-UP}($G,\alpha \cup \{(x,1-\epsilon)\})$
  \end{tabbing}
\end{minipage}
\caption{DLL with Clause Learning.}
\label{fig:dll-l-up}
\end{center}
\end{figure}

The \textsc{DLL-L-UP} algorithm as shown in \pref{fig:dll-l-up} does not
explicitly include unit propagation.  Rather, the use of unit propagation is
hidden in the test on line~2 of whether unit propagation can be used to
find a conflict graph.  In practice, of course, most algorithms set
variables by unit propagation as soon as possible and update
the implication graph each time a new unit variable is set.  The
algorithm as formulated in \pref{fig:dll-l-up} is more general, and thus
covers more possible implementations of \textsc{DLL-L-UP}, including
algorithms that may change the implication graph retroactively or may pick
among several conflict graphs depending on the details of how $F$~can be
falsified.  There is at
least one implemented clause learning algorithm that does this \cite{FMM:SAT04zChaff}.

As shown in \pref{fig:dll-l-up}, if $\rest F \alpha$ is false, then the
algorithm must return \texttt{UNSAT} (lines 2-6). 
Sometimes, however,
we use instead a ``non-greedy'' version of \textsc{DLL-L-UP}.  For the
non-greedy version it is optional for the algorithm to immediately return
\texttt{UNSAT} once $F$ has a conflict graph.
Thus the non-greedy \textsc{DLL-L-UP} algorithm can set
a branching variable (lines 7-11) even if $F$~has already been falsified
and even if there are unit clauses present.  
This non-greedy version of \textsc{DLL-L-UP} will
be used in the next section to simulate
regWRTI proofs.

The constructions of
Section~\ref{sec:regwrti-dll}
also imply that \textsc{DLL-L-UP} is p-equivalent
to the restriction of \textsc{DLL-L-UP} in which only series
decompositions are allowed.  That is to say, \textsc{DLL-L-UP}
with only series decompositions can simulate any run of
\textsc{DLL-L-UP} with at most polynomially many more recursive
calls.

% \textsc{DLL-L-UP} with restarts is an approach to describe the
% capabilities of modern, imperative SAT solvers like Zchaff or MiniSAT
% in a short and abstract way that is separated from implementation
% details.  If our definition does not match the implementations of DLL
% based SAT solvers then it would be interesting to discuss how single
% implementations defer from this schema to find a better abstract
% description of these SAT solvers that can be used for future
% theoretical considerations.

%%% Local Variables: 
%%% mode: latex
%%% TeX-master: "../main"
%%% End: 

\section{Equivalence of regWRTI and DLL-L-UP}\label{sec:regwrti-dll}

\subsection{regWRTI simulates DLL-L-UP}

We shall prove that regular WRTI proofs are equivalent to
non-greedy \hbox{\textsc{DLL-L-UP}} searches.  We start by showing that
every \textsc{DLL-L-UP} search can be converted into a
regWRTI proof.  As a first step, we prove that, for a
given series-parallel decomposition~$\mathcal H$ of a conflict graph, there
is a single regWRTI proof~$T$ such that every learnable clause of~$\mathcal H$
appears as an input-derived
clause in~$T$. Furthermore, $T$~is polynomial size;
in fact,
$T$ has size at most quadratic in the number of distinct variables
that appear in the conflict graph.

This theorem generalizes earlier, well-known results of Chang
\cite{Chang1970} and Beame et al.~\cite{BeameKautzSabharwal2004} that 
any individual learned clause can be derived by input resolution (or, more
specifically, that unit resolution is equivalent to input resolution).
The theorem states a similar fact about proving an entire
set of learnable clauses simultaneously.

\begin{thm} \label{the:regWRTIforLearnables}
Let $G$~be a conflict graph of size~$n$ for~$F$ under the assignment~$\alpha$.
Let $\mathcal H$ be a series-parallel decomposition for~$G$.
Then there is a regWRTI proof~$T$ of size~$\le n^2$
such that every learnable clause
of~$\mathcal H$ is an input-derived clause in~$T$. The final clause of~$T$
is equal to $\Cc{G}$.
Furthermore, $T$~uses as
resolution variables, only variables that are used as
nodes (possibly negated) in
$G\setminus \leafs{G}$.
\end{thm}

First we prove a lemma.  
Let the subconflict graphs $H_0\subset H_1\subset \cdots \subset H_k$
and $H_{0,0}\subset H_{0,1} \subset \cdots \subset H_{k-1,m_{k-1}}$
be as in the definition of series-parallel decomposition.  
\begin{lem} \label{lem:lemmaA}
\hspace*{1em}
\begin{enumerate}[\em(a)]
\item There is an input proof~$T_0$ from~$F$ which contains
every conflict clause $\CC {H_{0,j}}$, for $j=1,\ldots,m_0$.
Every resolution variable in~$T_0$
is a non-leaf node (possibly negated) in~$H_1$.
\item Suppose that $1\le i<k$ and $u$~is a literal in $\leafs{H_i}$.
Then there is an input proof~$T^u_i$ which contains every
(existing) induced clause $\ic{u}{H_{i,j}}$ for $j=1,\ldots,m_i$.
Every resolution variable in~$T^u_i$ is a non-leaf node (possibly negated)
in the subgraph $(H_{i+1})_u$ of~$H_{i+1}$ rooted at~$u$.
\end{enumerate}
\end{lem}

\proof
We prove part~a.\ of the lemma and then indicate the minor
modifications needed to prove part~b.
The construction of~$T_0$ proceeds by induction on~$j$ to build
proofs $T_{0,j}$; at the end, $T_0$~is set equal to~$T_{0,m_0}$.
Each proof $T_{0,j}$ ends with the clause $\CC{H_{0,j}}$ and contains
the earlier proof~$T_{0,j-1}$ as a subproof.
In addition, the only variables used as resolution variables in~$T_{0,j}$
are variables that are non-leaf nodes (possibly negated) in~$H_{0,j}$.

To prove the base case $j=1$, we must show that
$\CC{H_{0,1}}$ has an input proof~$T_{0,1}$.
Let the two immediate predecessors of~$\Box$ in~$G$ be the literals $x$
and~$\overline x$.
Define a clause~$C$ as follows.  If $x$ is not a leaf in~$H_{0,1}$, 
then we let $C = C_x$; recall that $C_x$~is the clause that contains
the literal~$x$ and the negations of literals that are immediate
predecessors of~$x$ in the conflict graph.  Otherwise,
since $H_{0,1}\not= H_0$, $\overline x$ is not a leaf in~$H_{0,1}$,
and we let
$C=C_{\overline x}$. 
By inspection,
$C$~has the
property that it contains only negations of literals that are in~$H_{0,1}$. 
For $l\in C$, define
the $\{0,1\}$-depth of~$l$ as the maximum length
of a path to~$\overline l$ from a leaf of~$H_{0,1}$.  If all literals in~$C$
have $\{0,1\}$-depth equal to zero, then $C = \CC{H_{0,1}}$, 
and $C$~certainly has an input proof from~$F$ (in fact, since $C=C_x$
or $C=C_{\overline x}$, we must have $C\in F$).

Suppose on the other hand, 
that $C$ is a subset of the nodes of~$H_{0,1}$ with
some literals of non-zero $\{0,1\}$-depth. 
Choose a literal~$l$ in~$C$
of maximum $\{0,1\}$-depth~$d$ and 
resolve $C$ with the clause $C_{\overline {l}}\in F$ to
obtain a new clause~$C^\prime$.  Since $C_{\overline {l}}\in F$,
the resolution step introducing~$C^\prime$ preserves the property of
having an input proof from~$F$. 
Furthermore, the new literals in~$C^\prime\setminus C$ 
have $\{0,1\}$-depth strictly less than~$d$. 
Redefine $C$~to be the just constructed clause~$C^\prime$.  If
this new $C$ is a subset of~$\CC{H_{0,1}}$ we are done constructing~$C$. 
Otherwise,
some literal in~$C$ has non-zero $\{0,1\}$-depth.  In this latter case,
we repeat the above construction to obtain a new~$C$, and continue
iterating this process
until we obtain~$C\subset \CC{H_{0,1}}$.

When the above construction is finished, $C$~is constructed as a clause
with a regular input proof~$T_{0,1}$ from~$F$ (the regularity follows by the
fact that variables introduced in~$C^\prime$ have $\{0,1\}$ depth less than
that of the resolved-upon variable).  Furthermore $C\subset \CC{H_{0,1}}$.  
In fact, $C = \CC{H_{0,1}}$ must hold, because there is a path, in~$H_{0,1}$,
from each leaf of~$H_{0,1}$ to~$\Box$.
That completes the proof of the $j=1$ base case.

For the induction step, with $j>1$,
the induction hypothesis is that we have constructed
an input proof~$T_{0,j}$ such that
$T_{0,j}$ contains all the clauses $\CC{H_{0,p}}$ for $1\le p \le j$ and
such that the final clause in~$T_{0,j}$ is the clause $\CC{H_{0,j}}$.
We are seeking to extend this input proof to an input proof
$T_{0,j+1}$ that ends with the
clause $\CC{H_{0,j+1}}$.  The construction of~$T_{0,j+1}$ proceeds exactly
like the construction above of~$T_{0,1}$, but now we start with 
the clause $C = \CC{H_{0,j}}$ (instead of $C=C_x$ or~$C_{\overline x}$),
and we update~$C$ by choosing the literal~${l}\in C$
of maximum $\{0,j+1\}$-depth
and resolving with~$C_{\overline {l}}$ to derive the next~$C$.
The rest of the construction of~$T_{0,j+1}$ is similar to
the previous argument.
For the regularity of the proof it is essential that $H_{0,j}$ is a
proper subconflict graph of $H_{0,j+1}$.
By inspection, any literal~$l$ used for resolution in the new
part of~$T_{0,j+1}$ is a non-leaf node in~$H_{0,j+1}$ and has a path
from~$l$ to some leaf node of~$H_{0,j}$.  Since $H_{0,j}$ is proper,
it follows that $l$~is not an inner node of~$H_{0,j}$ and thus is
not used as a resolution literal in~$T_{0,j}$.  Thus $H_{0,j+1}$ is
regular.
This completes the proof of part~a.

The proof for part~b.\ is very similar to the proof for part~a.
Fixing $i>0$, let $u$~be any literal in $\leafs{H_{i,0}}$.  We
need to prove, for $1\le j\le m_i$, there is an input proof~$T_{i,j}^u$
from~$F$
such that 
(a)~$T_{i,j}^u$~contains every existing induced clause $\ic u {H_{i,k}}$ for
$1\le k<j$, and (b)~$T_{i,j}^u$ ends with the
induced clause $\ic u {H_{i,j}}$,
and (c)~the resolution variables used in~$T^u_{i,j}$ are all non-leaf nodes
(possibly negated) of $V_{(H_{i,j})_u}$.  The proof is by induction on~$j$.
One starts with the clause $C = C_u$.  The main step of the construction
of $T^u_{i,j+1}$ from~$T^u_{i,j}$ is to find the literal $v\not=u$ in~$C$
of maximum $\{i,j\}$-depth, and resolve $C$ with~$C_{\overline v}$
to obtain the next~$C$.  This process proceeds iteratively
exactly like the construction
used for part~a.
This completes the proof of \pref{lem:lemmaA}.
\qed

We now can prove \pref{the:regWRTIforLearnables}.  \pref{lem:lemmaA}
constructed separate regular input resolution
proofs $T_{0,m_0}=T_0$ and~$T_{i,m_i}^u=T_i^u$ that included
all the learnable clauses of~$\mathcal H$.  To complete the proof
of \pref{the:regWRTIforLearnables}, we combine all these proofs
into one single regWRTI proof.  For this, we construct
proofs $T^*_i$ of the clause $\CC {H_i}$.  $T^*_1$~is just~$T_{0}$. 
The proof~$T^*_{i+1}$ is constructed from~$T^*_i$ by
successively resolving the final clause of~$T^*_i$ with the final clauses
of the proofs~$T^u_i$,
using each $u \in \leafs {H_i}\setminus \leafs{H_{i+1}}$ as
a resolution variable, taking
the~$u$'s in order of increasing $\{i,m_i\}$-depth to preserve
regularity.  
Letting $T = T^*_k$,
it is clear
that $T^*_k$~contains all the clauses from~$\Cc{\mathcal H}$,
and, by construction, $T^*_k$~is regular.
%% MAYBE THIS REGULARITY SHOULD GET PROVED.  

To bound the size of~$T$, note that any
regular input proof~$S$ has size $2r+1$ where $r$ is
the number of distinct variables used as resolution variables in~$S$.
Since $T$ is regular, and is formed by combining the regular
input proofs $T_0$, $T^u_i$ in a linear fashion, the total size
of~$T$ is less than $ n + \sum_{k=0}^{n-1}(2k+1) = n^2+1$.
This completes the proof of \pref{the:regWRTIforLearnables}.
\hfill $\Box$

Note that, since the final clause of~$T$ contains only literals
from $\leafs G$, $T$~does not use any variable that occurs in its final
clause as a resolution variable.

\medskip

We can now prove the first main result of this section, namely, that
regWRTI proofs polynomially simulate \textsc{DLL-L-UP} search trees.

\begin{thm}\label{the:rWRTIsimDLL}
Suppose that $F$~is an unsatisfiable set of clauses and that there is
an execution of a
(possibly non-greedy)
\textsc{DLL-L-UP} search algorithm on input~$F$ that outputs \texttt{UNSAT}
with $s$~recursive calls.  Then there is a regWRTI refutation of~$F$
of size at most $s\cdot n^2$ where $n = |\var(F) |$.
\end{thm}

\proof
Let $S$ be the search tree associated with the \textsc{DLL-L-UP}
algorithm's execution.
We order~$S$ so that the \textsc{DLL-L-UP} algorithm effectively 
traverses~$S$ in
a depth-first, left-to-right order.  We transform~$S$ into
a regWRTI proof tree~$T$ as follows.  The tree~$T$ contains a copy
of~$S$, but adds subproofs at the leaves of~$S$ (these subproofs will
be derivations of learned clauses).  For each internal node in~$S$,
if the corresponding branching variable was~$x$ and was first set
to the value~$x^\epsilon$, then the corresponding node in~$T$ is
labeled with $x$ as the resolution variable, and its left incoming
edge is labeled with~$x^{\epsilon}$ and its right incoming edge
is labeled with~$x^{1-\epsilon}$.  
For each node~$u$ in~$S$, let $\alpha_u$~be
the assignment at that node that is held by the 
\textsc{DLL-L-UP} algorithm upon
reaching that node.  
By construction, $\alpha_u$~is equivalently defined as the assignment
that has $\alpha_u(l) = 1$ for literal~$l$
that labels an edge
on the path (in~$T$) between~$u$ and the root of~$T$. 

For a node~$u$ that is a leaf of~$S$, the \textsc{DLL-L-UP} algorithm chooses
a conflict graph~$G_u$ with 
a series-parallel decomposition~${\mathcal H}_u$ such that every
leaf node~$l$ of~$G_u$ is a literal set to
true by~$\alpha_u$.  Also, let~$F_u$ be the
set~$F$ of original clauses augmented with all clauses learned
by the \textsc{DLL-L-UP} algorithm before reaching node~$u$.   
By \pref{the:regWRTIforLearnables},
there is a proof~$T_u$ from the clauses~$F_u$ such that
every learnable clause of~${\mathcal H}_u$ appears in $T_u$ as in
input-derived clause.  Hence, of course, every clause learned at~$u$ by the
\textsc{DLL-L-UP} algorithm appears in~$T_u$ as an input-derived clause.
The leaf node~$u$ of~$S$ is then
replaced by the proof~$T_u$ in~$T$. 
Note that by \pref{the:regWRTIforLearnables} and
the definition of conflict graphs, the final clause~$C_u$ of~$T_u$
is a clause that contains only literals falsified by~$\alpha_u$.  

So far, we have defined the clauses~$C_u$ that label nodes~$u$ in~$T$
only for leaf nodes~$u$.  For internal nodes~$u$, we define $C_u$~inductively
by letting $v$ and~$w$ be the immediate predecessors of~$u$ in~$T$ and
defining $C_u$~to be the clause obtained by (w-)resolution 
from the clauses $C_v$ and~$C_w$ with respect to the branching
variable~$x$ that was picked at node~$u$ by the \textsc{DLL-L-UP}
algorithm.  Clearly, using induction from the leaves of~$S$,
the clause~$C_u$ contains only variables that are falsified by the
assignment~$\alpha_u$.  This makes $T$ a regWRTI
proof.  

Let $r$~be the root node of~$S$.  Since $\alpha_r$~is the empty assignment,
the clause~$C_r$
must equal the empty clause~$\Box$.  Thus $T$~is a regWRTI refutation
of~$F$ and \pref{the:rWRTIsimDLL} is proved.
\qed

Since DLL clause learning based on first cuts has been shown
to give exponentially shorter proofs than 
regular resolution~\cite{BeameKautzSabharwal2004},
and since
\pref{the:rWRTIsimDLL} states that regWRTI can simulate DLL
search algorithms (including ones that learn first cut clauses),
we have proved that regRD does not simulate regWRTI:
\begin{thm}\label{the:regRDnosimregWRTI}
$\text{regRD} <  \text{regWRTI}$.
\end{thm}
Hoffmann \cite{Hoffmann2007} gave a
direct proof of \pref{the:regRDnosimregWRTI} based on the variable
extensions described below in \pref{sec:varexp}.

\subsection{DLL-L-UP simulates regWRTI}

We next show that the non-greedy \textsc{DLL-L-UP} search procedure can simulate
any regWRTI proof~$T$.  The intuition is that we split~$T$ into two
parts: the \emph{input parts} are the subtrees of~$T$ that contain
only input-derived clauses.  The \emph{interior part} of~$T$ is the rest of~$T$.
The interior part will be simulated by a \textsc{DLL-L-UP} search procedure
that traverses the tree~$T$ and at each node, chooses the resolution 
variable as the branching variable and sets the branching variable
according to the label on the left incoming edge.  In this way, the
tree~$T$ is traversed in a depth-first, left-to-right order.  The
input parts of~$T$ are not traversed however.  Once an input-derived clause
is reached, the \textsc{DLL-L-UP} search learns all the clauses in
that input subproof and backtracks returning \texttt{UNSAT}.

The heart of the procedure is how a conflict graph and corresponding
series-parallel decomposition can be picked so as to make all the
clauses in a given input subproof learnable.  This is the content of
the next lemma.
\begin{lem}\label{lem:lemmaB}
Let $T$~be a regular input proof of~$C$ from a set of clauses~$F$.
Suppose that $\alpha$ falsifies~$C$, that is, $\rest C \alpha = 0$.
Further suppose no variable in~$C$ is used as a resolution variable
in~$T$.
Then there is a conflict graph~$G$ for~$F$ under~$\alpha$ and
a series decomposition~$\mathcal H$ for~$G$ such that the set of learnable
clauses of~${\mathcal H}$ is equal to the set of input-derived clauses of~$T$.
\end{lem}
Recall that a series decomposition just means a series-parallel decomposition
with a trivial parallel part, i.e, $k=1$ in the definition of
series-parallel decompositions.
\proof
Without loss of generality, $F$~is just the set of initial
clauses of~$T$.  Let the input proof~$T$ contain clauses $C_{m+1}=C,
C_{m}, \ldots,C_1, D_{m},\ldots,D_1$ as illustrated in 
\pref{fig:regRTI} with $m=4$.  Each $C_{i+1}$~is inferred from $C_{i}$
and~$D_{i}$ by resolution on~$l_{i}$, where
$\overline{l_i}\in C_i$ and $l_i \in D_i$.
For each~$i$, we have
$D_i = \{l_i\} \cup D^\prime_i$, where $ D^\prime_i\subseteq C_{i+1}$.
Likewise,
$C_i = \{\overline{l_i}\} \cup C^\prime_i$,
where $ C^\prime_i\subseteq C_{i+1}$.
\begin{figure}
\begin{center}
\psset{unit=1cm} % Use this to scale the pspicture
\begin{pspicture}(-6,-0.2)(3,4)
\pscircle*(-4,4){0.07}
\pscircle*(-2,4){0.07}
\pscircle*(-3,3){0.07}
\pscircle*(-1,3){0.07}
\pscircle*(-2,2){0.07}
\pscircle*(0,2){0.07}
\pscircle*(-1,1){0.07}
\pscircle*(1,1){0.07}
\pscircle*(0,0){0.07}
\psline(-4,4)(0,0)
\psline(-2,4)(-3,3)
\psline(-1,3)(-2,2)
\psline(0,2)(-1,1)
\psline(1,1)(0,0)
\uput[-45](0.5,0.5){$\overline{l_4}$}
\uput[-45](-0.5,1.5){$\overline{l_3}$}
\uput[-45](-1.5,2.5){$\overline{l_2}$}
\uput[-45](-2.5,3.5){$\overline{l_1}$}
\uput[-135](-0.5,0.5){$l_4$}
\uput[-135](-1.5,1.5){$l_3$}
\uput[-135](-2.5,2.5){$l_2$}
\uput[-135](-3.5,3.5){$l_1$}
\uput[45](1,1){$D_4$}
\uput[45](0,2){$D_3$}
\uput[45](-1,3){$D_2$}
\uput[45](-2,4){$D_1$}
\uput[225](-1,1){$C_4$}
\uput[225](-2,2){$C_3$}
\uput[225](-3,3){$C_2$}
\uput[135](-4,4){$C_1$}
\uput[-90](0,0){$C_{5}=C$}
\end{pspicture}
\end{center}
\caption{A regular input proof of~$C$.  Edges are labeled $l_i$ or
$\overline{l_i}$.  The $C_i$'s and $D_i$'s are clauses.}
\label{fig:regRTI}
\end{figure}

As illustrated in Figure~\ref{fig:conflictDecomp},
we construct conflict graphs
$H_{0,0} = \{\Box,l_1,\overline{l_1}\} \subset H_{0,1}
\subset \cdots \subset H_{0,m} =G$ which form a series decomposition
of~$G$.  $H_{0,i}$~will be a conflict graph
from the set of clauses $\{C_1,D_1,\ldots,D_i\}$ under~$\alpha_i$ where
$\alpha_i$~is the assignment that falsifies all the literals in~$C_{i+1}$. 
Indeed, the leaves of~$H_{0,i}$ are precisely the negations
of literals in~$C_{i+1}$. 
For $i>0$, the
non-leaf nodes of $H_{0,i}$ are $\overline{l_1}$ and $l_1,\ldots,l_i$.  The
predecessors of~$\overline{l_1}$ are defined to be the literals~$u$
with $\overline{u} \in C_1^\prime$, that is $C_{\overline{l_1}} = C_1$.
Likewise, the predecessors of~$l_i$ are
the literals~$u$ with $\overline{u} \in D_i^\prime$ so that 
$C_{l_i} = D_i$.

To start with, we define $H_{0,0}$ to equal $\{\Box, l_1, \overline l_1\}$.
Let $H_{0,i}$ be already
constructed.  Then we have $\overline{l}_{i+1} \in C_{i+1}$ since
$C_{i+2}$ is inferred by 
resolution on~$l_{i+1}$ from~$C_{i+1}$.
It follows that $\alpha_i(l_{i+1}) = 1$ and that $l_{i+1}$~is
a leaf in~$H_{0,i}$.  We obtain~$H_{0,i+1}$ from~$H_{0,i}$ by adding the
predecessors of~$l_{i+1}$ (i.e., the literals~$u$ with $\overline{u} \in
D_{i+1}^\prime $) to~$H_{0,i}$.  The leaves of~$H_{0,i+1}$ are now exactly
the negations of the literals in the clause~$C_{i+2}^\prime$.  Finally
the graph $H_{0,m} = G$ and the series decomposition $\mathcal{H}$ defined
by the graphs $H_{0,i}$ is as wanted.  This completes the proof of
\pref{lem:lemmaB}.
\qed

\begin{figure}
\begin{center}
\psset{yunit=1.2cm}
\psset{xunit=0.8cm}
\begin{pspicture}
\begin{pspicture}(-6,-0.2)(7,8)
\rput(0,0){$\Box$}
\pnode(0,0){BOX}
\pscircle(0,0){0.4cm}
\rput(0,6){$l_4$}
\pnode(0,6){L4}
\pscircle(0,6){0.4cm}
\rput(1.5,4.5){$l_3$}
\pnode(1.5,4.5){L3}
\pscircle(1.5,4.5){0.4cm}
\rput(3.0,3.0){$l_2$}
\pnode(3.0,3.0){L2}
\pscircle(3.0,3.0){0.4cm}
\rput(4.5,1.5){$l_1$}
\pnode(4.5,1.5){L1}
\pscircle(4.5,1.5){0.4cm}
\rput(-4.5,1.5){$\overline{l_1}$}
\pnode(-4.5,1.5){L1neg}
\pscircle(-4.5,1.5){0.4cm}
\psset{nodesep=0.4cm}
\ncline{->}{L4}{L3}
\ncline{->}{L3}{L2}
\ncline{->}{L2}{L1}
\ncline{->}{L1}{BOX}
\ncline{->}{L4}{L1neg}
\ncline{->}{L3}{L1neg}
\ncline{->}{L2}{L1neg}
\ncline{->}{L1neg}{BOX}
\rput(5.5,3.0){$D^{\prime\prime}_1$}
\rput(4.2,4.7){$D^{\prime\prime}_2$}
\rput(2.7,6.2){$D^{\prime\prime}_3$}
\rput(0.0,7.5){$D^{\prime\prime}_4$}
\rput(-5.2,3.2){$C^{\prime\prime}_1$}
\pnode(5.5,3.0){D1}
\pnode(4.2,4.7){D2}
\pnode(2.7,6.2){D3}
\pnode(0.0,7.5){D4}
\pnode(-5.2,3.2){C1}
\ncline[doubleline=true]{->}{D1}{L1}
\ncline[doubleline=true]{->}{D2}{L2}
\ncline[doubleline=true]{->}{D3}{L3}
\ncline[doubleline=true]{->}{D4}{L4}
\ncline[doubleline=true]{->}{C1}{L1neg}
\psset{arcangle=-25}
\ncarc{->}{L4}{L2}
\ncarc{->}{L3}{L1}
\psset{arcangle=-35}
\ncarc{->}{L4}{L1}
\psset{linestyle=dotted,linewidth=1.5pt}
\psline(-6.5,2.25)(6.5,2.25)
\psline(-6.1,3.75)(6.1,3.75)
\psline(-5.7,5.25)(5.7,5.25)
\psline(-5.3,6.75)(5.3,6.75)
\psline(-4.9,8.25)(4.9,8.25)
\uput[0](6.5,2.25){$H_{0,0}$}
\uput[0](6.1,3.75){$H_{0,1}$}
\uput[0](5.7,5.25){$H_{0,2}$}
\uput[0](5.3,6.75){$H_{0,3}$}
\uput[0](4.9,8.25){$H_{0,4}$}
% \psset{linestyle=dotted,linewidth=1.2pt}
% \pscurve(-5.3,1.5)(-4.5,2.2)(0.0,2.5)(4.7,2.2)(5.3,1.5)
% \pscurve(-5.5,1.5)(-4.5,2.5)(0.0,3.8)(3.5,3.7)(4.0,3.3)(5.5,1.5)
% \pscurve(-5.7,1.5)(-4.5,2.7)(0.0,5.0)(1.9,5.2)(3.8,3.7)(5.7,1.5)
% \pscurve(-5.9,1.5)(-4.5,2.9)(0.0,6.8)(3.8,4.0)(5.9,1.5)
\end{pspicture}
\end{center}
\caption{A conflict graph and a series decomposition.  The solid lines
and arcs
indicate edges that may or may not be present.  
The notations $C^{\prime\prime}_1$
and $D^{\prime\prime}_i$ indicate zero or more literals, and
the double lines indicate an edge from each literal in the set.
The dashed lines indicate
cuts, and thereby the sets $H_{0,i}$ in the
series decomposition. Namely, 
the set~$H_{0,i}$ contains the nodes below the corresponding
dotted line.}
\label{fig:conflictDecomp}
\end{figure}

We can now finish the proof that \textsc{DLL-L-UP} simulates
regWRTI.

\begin{thm}\label{the:DLLsimrWRTI}
Suppose that $F$ has a regWRTI proof of size~$s$.  Then there is
an execution of the non-greedy {\rm \textsc{DLL-L-UP}}
algorithm with the input \hbox{\rm ($F,\varnothing$)}
that makes $<s$~recursive calls.
\end{thm}

\proof
Let $T$~be a regWRTI refutation of~$F$.  The \textsc{DLL-L-UP} algorithm
works by traversing the proof tree~$T$ in a depth-first, left-to-right order.
At each non-input-derived node~$u$ of~$T$, labeled with
a clause~$C$, the resolution variable for that clause
is chosen as the branching variable~$x$, and the variable~$x$ is
assigned the value 1 or~0, corresponding to the label on the
edges coming into~$u$.  By part~b.\ of \pref{the:regWprops},
the clause~$C$ is
falsified by the assignment~$\alpha$.  At each input-derived node of~$T$,
the \textsc{DLL-L-UP} algorithm learns the clauses in the input subproof
above~$u$ by using the conflict graph and series decomposition given
by \pref{lem:lemmaB}.  Since the \textsc{DLL-L-UP} search cannot find
a satisfying assignment, it must terminate after traversing the (non-input)
nodes in the regWRTI refutation tree.  The number of recursive calls will
equal twice the number of non-input-derived nodes of~$T$,
which is less than~$s$.
\qed

%%% Local Variables: 
%%% mode: latex
%%% TeX-master: "../main"
%%% End: 

\section{Generalized DLL with clause learning}\label{sec:dll-learn}

\subsection{The algorithm DLL-Learn}
This section presents a new formulation of DLL with learning 
called \textsc{DLL-Learn}.  This algorithm differs from
\textsc{DLL-L-UP} in two important ways.  First, unit propagation is
no longer used explicitly (although it can be simulated).  Second,
the \textsc{DLL-Learn} algorithm uses more information that arises
during the DLL search process, namely, it can infer clauses
by resolution at each node in the search tree.  This makes it
possible for \textsc{DLL-Learn} to simulate regular resolution trees with
full lemmas; more
specifically, \textsc{DLL-Learn} is equivalent to
regWRTL.

The \DLLlearn{} algorithm is very similar to the pool resolution
system introduced by Van Gelder~\cite{VanGelder2005}.  Furthermore,
our Theorem~\ref{the:DLL-Learn=regwRTL}
is similar to results obtained by Van Gelder
for pool resolution.
Our constructions differ mostly 
in that we use w-resolution in place
of the degenerate resolution inference of Van Gelder~\cite{VanGelder2005}. 
Loosely speaking,
Van Gelder's degenerate resolution inference is a method of allowing
resolution to operate on any two clauses without any weakening.  Conversely, 
our w-resolution is a method for allowing resolution to operate on
any two clauses, but with the maximum reasonable amount of weakening.

The idea of \textsc{DLL-Learn} is to extend DLL
so that it can learn a new clause~$C$ at each node in the
search tree.  As usual, the new clause will
satisfy $F \equiv F\cup\{C\}$.
At leaves, \textsc{DLL-Learn} does not learn a
new clause, but marks a preexisting falsified clause as ``new''.
At internal nodes, after branching on a variable~$x$ and
making two recursive calls, the \textsc{DLL-Learn} algorithm can
use w-resolution to infer a new clause, 
$C_{DLL(F,\alpha)}$, from the two identified new clauses, $C_0$ and~$C_1$
returned
by the recursive calls.  
Since $x$ does not have to occur in $\var(C_0)$ and~$\var(C_1)$, 
$C$~is obtained by a w-resolution instead of resolution.

The \textsc{DLL-Learn}
algorithm shown in \pref{fig:dll-learn}
uses non-greedy detection of contradictions.
Namely, the ``{\tt optionally do}'' on line~2 of \pref{fig:dll-learn}
allows the algorithm to
continue to branch on variables even if the formula is already
unsatisfied.  
This feature is 
needed for a direct proof of \pref{the:DLL-Learn=regwRTL}.
In addition, it could be helpful in an implementation of the
algorithm: Think of a call of \textsc{DLL}$(F,\alpha)$ such that
$\rest{F}{\alpha} = 0$ and suppose that all of the falsified clauses
$C \in F$ are very large and thus undesirable to learn.
It might, for example, be the case that $\rest{F}{\alpha}$ contains
two conflicting unit clauses $\rest{C_0}{\alpha}=\{x\}$ and
$\rest{C_1}{\alpha}= \{\neg x\}$, where $C_0$ and~$C_1$ are small.
In that case, it could be better to branch on the
variable~$x$ and to learn the resolvent of $C_0$ and~$C_1$.

There is one situation where it is not optional to
execute lines 3-4;
namely, if $\alpha$~is a total assignment and
has assigned values to all variables, then the algorithm must do
lines 3-4.

Note that it is possible to remove $C_0$ and~$C_1$ from $F$ in line~13
if they were previously learned.  Additionally, in an implementation
of \textsc{DLL-Learn} it could be helpful to tag~$C_i$ as the new
clause in~$H$ in line~13 if $C_i \subseteq C$ for an $i\in\{0,1\}$ instead of
learning~$C$ --- this would be essentially equivalent to using
Van Gelder's degenerate resolution instead of
w-resolution.

\begin{figure}[htbp]
\begin{center}
\begin{minipage}{1.0\linewidth}
 \tt \small
  \begin{tabbing}
  123\=123455\=12345\=12345\=12345\=12345\=12345\=12345\=12345\=12345\=12345\=12345 \kill
  \>{\sc DLL-Learn}($F,\alpha$)\\
  \>1\>if $\rest{F}{\alpha} = 1$ then return ($F,\alpha$) \\
  \>2\>if $\rest{F}{\alpha} = 0$ then optionally do \>\>\>\>\>\>\>\>\> \\
  \>3\>\>tag a $C \in F$ with $\rest{C}{\alpha}=0$ as the new clause\\
  \>4\>\>return ($F,$\hspace{0.2em}UNSAT)\\
  \>5\>choose $x \in \var(F) \setminus \dom(\alpha)$ and a value $\epsilon \in \{0,1\}$\\
  \>6\>($G,\beta$)$\leftarrow${\sc DLL-Learn}($F,\alpha \cup \{(x,\epsilon)\})$\\
  \>7\>if $\beta \neq $ UNSAT then return ($G,\beta$)\\
  \>8\>($H,\gamma$)$\leftarrow${\sc DLL-Learn}($G,\alpha \cup \{(x,1-\epsilon)\})$\\
  \>9\>if $\gamma \neq$ UNSAT then return ($H,\gamma$)\\
  \>10\>select the new $C_0 \in G$ and the new $C_1 \in H$\\
  \>11\>$C \leftarrow (C_0 - \{x^{1-\epsilon}\}) \cup (C_1 - \{x^\epsilon\})$\\
  \>12\>$H \leftarrow H \cup \{C\}$ \>\>\>\>\>\>\>\>\> -- {\sl learn a clause}\\
  \>13\>tag $C$ as the new clause in~$H$. \\
  \>14\>return ($H,$\hspace{0.2em}UNSAT)
  \end{tabbing}
\end{minipage}
\caption{DLL with a generalized learning.}
\label{fig:dll-learn}
\end{center}
\end{figure}

It is easy to verify that, at any point in the \textsc{DLL-Learn}
algorithm, when a clause~$C$ is tagged as new, then $\rest C \alpha = 0$.

There is a straightforward, and direct, translation between executions
of the \textsc{DLL-Learn} search algorithm on input $(F,\varnothing)$ and
regWRTL proofs of~$F$.  An execution of \textsc{DLL-Learn}($F,\varnothing$)
can be
viewed as traversing a tree in depth-first, left-to-right order.  If there 
are $s-1$ recursive calls to \textsc{DLL-Learn}, the tree has $s$~nodes.
Each node of the search tree is labeled with the clause tagged in the
corresponding call to \textsc{DLL-Learn}.  Thus, leaves of the
tree are labeled with clauses that either are from~$F$ or were learned
earlier in the tree.  The clause on an internal node of the tree
is inferred from the clauses on the two
children using w-resolution with respect to the branching variable.
Finally, the clause~$C$ labeling the root node,
where $\alpha = \varnothing$, must
be the empty clause, since $\alpha$~must falsify~$C$.
In this way the search algorithm describes precisely a regWRTL
proof tree.  Conversely, any regWRTL refutation of~$F$ corresponds exactly
to an execution of the \textsc{DLL-Learn}($F,\varnothing$). 

This translation between \textsc{DLL-Learn} and regWRTI proof trees
gives the following theorem.

\begin{thm}\label{the:DLL-Learn=regwRTL}
Let $F$~be a set of clauses.
There exists a regWRTL refutation of~$F$ of size~$s$
if and only if there is an execution of
\textsc{DLL-Learn}$(F,\varnothing)$ that performs exactly $s-1$
recursive calls.
\end{thm}

It follows as a corollary of 
Theorems \ref{the:hierarchy} and~\ref{the:DLL-Learn=regwRTL}
that \textsc{DLL-Learn} can polynomially
simulate \textsc{DLL-L-UP}.

%%% Local Variables: 
%%% mode: latex
%%% TeX-master: "../main"
%%% End: 

\section{Variable Extensions}\label{sec:varexp}

This section introduces the notion of a \emph{variable extension} of a CNF
formula.  A variable extension augments a set~$F$ of clauses with additional
clauses such that modified formula $\ve F$~is satisfiable if and only if $F$
is satisfiable.
Variable extensions will be used to prove that regWRTI
proofs can simulate resolution dags, in the sense
that if there is an RD refutation of~$F$, then there is a 
polynomial size regWRTI refutation of~$\ve F$.
Hence,
\textsc{DLL-Learn}
and the non-greedy version of \textsc{DLL-L-UP}
can simulate full (non-regular) resolution in the same sense.

Our definition of
variable extensions is inspired by the proof trace extensions 
of Beame et al.~\cite{BeameKautzSabharwal2004} that were used to separate
DLL with clause learning from regular resolution dags.
A similar construction was used by Hertel et~al.~\cite{BHPvG:clauselearn}
to show that pool resolution can simulate full resolution.
Our results strengthen and extend the prior results by applying
directly to regWRTI proofs.
More importantly,
in contrast to proof trace extensions, variable extensions do
not depend on the size of a (possibly unknown) resolution proof but only
on the number of variables in the formula.

\begin{defi}
Let $F$ be a set of clauses
and $|\var(F)|=n$.
The set of \emph{extension variables} of~$F$ is $\ev{F} = \{q,p_1,
\ldots, p_n\}$, where $q$ and~$p_i$ are new variables.
The \emph{variable extension} of~$F$ is the set of clauses 
\[
\ve{F} ~=~ F
   \cup \big\{ \{q, \bar l\}:l \in C \in F\big\}
   \cup \big\{\{p_1,p_2, \ldots, p_n \}\big\}.
\]
\end{defi}
Obviously $\ve F$ is satisfiable if and only if~$F$ is.  Furthermore,
$|\ve F| = O(|F|)$.

Suppose that $G$ is a resolution dag (RD) proof from~$F$.  
We can reexpress~$G$ as a sequence of (derived) clauses $C_1,C_2,\ldots, C_t$
which has the following properties: (a)~$C_t$~is the final
clause of~$G$,
and (b)~each $C_i$ is inferred by resolution from two clauses $D$ and~$E$, 
where each of $D$ and~$E$
either are in~$F$ or
appear earlier in the sequence as $C_j$ with $j<i$.  Basically, the sequence
is an ordinary resolution refutation, but with the clauses from~$F$ omitted.

\begin{lem}\label{lem:pte_trick_helper}
Suppose that $D,E\vdash_x C$.  Then, there is
an input resolution proof tree~$T_C$ of the clause~$\{q\}$ from
$\ve F \cup \{D,E\}$ such that $C$~appears in~$T_C$ and such that
$|T_C| = 2\cdot |C|+3$.
\end{lem}
\proof
The proof~$T_C$ starts by resolving $D$ and~$E$ to yield~$C$.  It
then resolves successively with the clauses $\{q,\overline l\}$, 
for $l\in C$, to derive~$\{q\}$.
\qed

\begin{thm}\label{the:pte_trick}
Let $F$ be a set of clauses, $n = |\var(F)|$, and let $C$~be a clause.
Suppose that $G$ is a resolution dag proof of~$C$ from~$F$ of size~$s$.
Then, there is a regWRTI proof~$T$ of~$C$ from~$\ve F$
of size $\le 2s\cdot(d+2)+1$ where $d = \max \{ |D| : D\in G \}\le n$.
\end{thm}

\proof
Let $C_1,\ldots, C_t$ be a sequence of the derived clauses in~$G$ as above.  
Without loss of generality, $t< 2^n$ since $F$~also has a regular resolution
tree refutation, and this has depth at most~$n$, and thus has $<2^n$
internal nodes.
Let $T^\prime$~be a binary tree with $t$~leaves and 
of height~$h = \lceil \log_2 t \rceil \le n$.  For each 
node~$u$ in~$T^\prime$, let $l(u)$~be the level of~$u$ in ~$T^\prime$, namely,
the number of edges between $u$ and the root.  
Label $u$ with the variable~$p_{l(u)}$.  Also, label every node~$u$ in~$T^\prime$
with the clause~$\{q\}$.  $T^\prime$ will form the middle part of
a regWRTI proof: 
each clause $\{q\}$ at level $i$ is inferred by w-resolution from
its two children clauses (also equal to~$\{q\}$) with respect to the
variable~$p_i$.

Now, we expand $T^\prime$ into a 
regWRTI proof tree~$T^{\prime\prime}$.  For this, for $1\le i\le t$,
we replace the
$i$-th leaf of~$T^\prime$ with a new subproof~$T_{C_i}$ defined as follows.
Letting $C_i$ be as above, let $D_i$ and~$E_i$ be the
two clauses from which $C_i$~is inferred in~$G$.
Then replace $i$-th leaf of~$T^\prime$ by the input proof~$T_{C_i}$ from
\pref{lem:pte_trick_helper} which contains~$C_i$ and ends with the
clause~$\{q\}$.  Note that each of $D_i$ and~$E_i$ either is in~$F$ or appeared
as an input clause in a proof, $T_{D_i}$ or~$T_{E_i}$,
inserted at an earlier leaf of~$T^\prime$.  Therefore $T^{\prime\prime}$ is
a valid regWRTI proof of~$\{q\}$ from~$\ve F$.
Since there are at most $s-1$ internal nodes in~$T^\prime$ and each
$T_{C_i}$ has size $\le 2d+3$,
$T^{\prime\prime}$ has size
at most $(s-1) + s\cdot(2d+3)$.

Finally, we form a regWRTI proof of~$C$ by modifying~$T^{\prime\prime}$ by 
adding a new root labeled with the clause~$C$ and the resolution
variable~$q$.  Let the
left child of this new root be the root of~$T^{\prime\prime}$,
and let the right child be a new node labeled also with~$C$.  
(This is permissible since $C$~is input-derived in~$T^{\prime\prime}$.)
Label the left edge coming to the new root with the literal~$\overline q$,
and the right edge with the literal~$q$.   This makes $C$ inferred from
$\{q\}$ and~$C$ by w-resolution with respect to~$q$.
$T$~is a valid regWRTI of size at most $s+1+s\cdot(2d+3) = 2s\cdot(d+2)+1$.
\qed

Since \textsc{DLL-L-UP} and \textsc{DLL-Learn} simulate
regWRTI, \pref{the:pte_trick}
implies that these two systems p-simulate full resolution by the
use of variable extensions:

\begin{cor}
Suppose that $F$ has a resolution dag refutation of size~$s$.  Then both
\textsc{DLL-L-UP} and \textsc{DLL-Learn}, when
given $\ve F$ as input, have executions that return
\texttt{UNSAT} after at most $p(s)$ recursive calls, for some
polynomial~$p$.
\end{cor}

We now consider some issues about ``naturalness'' of proofs  based
on resolution with lemmas. 
Beame et al.~\cite{BeameKautzSabharwal2004} defined a refutation system
to be natural provided that, whenever $F$ has a refutation of size~$s$,
then $\rest F \alpha$ has a refutation of size at most~$s$.  We need
a somewhat relaxed version of this notion:
\begin{defi}
Let $\mathcal R$ be a refutation system for sets of clauses.
The system~$\mathcal R$ is {\em p-natural} provided, there is a polynomial~$p(s)$,
such that, whenever a set~$F$
has an $\mathcal R$-refutation of size~$s$, and $\alpha$~is a
restriction, then $\rest F \alpha$ has an $\mathcal R$-refutation
of size $\le p(s)$.
\end{defi}

The next proposition is well-known.

\begin{prop}
Resolution dags (RD) and regular resolution dags (regRD) are
natural proof systems.
\end{prop}

As a corollary to Theorem~\ref{the:pte_trick} we obtain the following
theorem.

\begin{thm}\label{the:equivNatural}
\hspace*{1em}
\begin{enumerate}[\em(a)]
\item regWRTI is equivalent to RD if and only if
regWRTI is p-natural.
\item regWRTL is equivalent to RD if and only if
regWRTL is p-natural.
\end{enumerate}
\end{thm}
\proof
Suppose that $\text{regWRTI}\equiv \text{RD}$.  Then, since RD is
natural, we have immediately that regWRTI is p-natural.

Conversely, suppose that regWRTI is p-natural.  By \pref{the:hierarchy},
RD p-\penalty10000simulates regWRTI.  So it suffices to
prove that regWRTI p-simulates RD.  Let $F$~have an
RD refutation of size~$s$.  By \pref{the:pte_trick}, $\ve F$
has a regWRTI proof of size~$2s(s+2)+1$. 
Let $\alpha$~be the assignment that assigns the value~$1$ to each of
the extension variables $q$ and $p_1,\ldots,p_n$.
Since $\rest {\ve F} \alpha$ is~$F$ and
since regWRTI is
p-natural, $F$~has a regWRTI proof of size at most $p(2s(s+2)+1)$.  This
proves that regWRTI p-simulates RD, and completes the proof of~a.

The proof of {b.} is similar.
\qed

Theorem~\ref{the:equivNatural} is stated for the equivalence of
systems with RD.  It could also be stated for {\em p-equivalent} but then
one needs an ``effective'' version of p-natural, where the
$\mathcal R$-refutation of~$\rest F \alpha$
is computable in
polynomial time from $\alpha$ and a $\mathcal R$-refutation of~$F$.

%%% Local Variables: 
%%% mode: latex
%%% TeX-master: "../main"
%%% End: 

\section{A Lower Bound for RTLW with short lemmas}
\label{sec:smlem}

In this section we prove a lower bound showing that learning only
short clauses does not help a DLL algorithm for certain hard formulas.
The proof system corresponding to DLL algorithms with learning
restricted to clauses of length $k$ is, according to
\pref{sec:regwrti-dll}, regWRTI with the additional restriction that
every used lemma is a clause of length at most $k$. We prove a lower
bound for a stronger proof system that allows arbitrary lemmas instead of just
input lemmas, drops the regularity restriction, and uses the general
weakening rule instead of just w-resolution, i.e., RTLW as defined
in \pref{sec:wrtl}.  We define RTLW($k$) to be the restriction of RTLW
in which every lemma used, i.e., every leaf label that does not occur
in the initial formula, is of size at most $k$.

The hard example formulas we prove the lower bound for are the
well-known Pigeonhole Principle formulas.  This principle states that
there can be no 1-to-1 mapping from a set of size $n+1$ into a set of
size $n$.  In propositional logic, the negation of this principle
gives rise to an unsatisfiable set of clauses $PHP_n$ in the variables
$x_{i,j}$ for $1 \leq i\leq n+1$ and $1 \leq j\leq n$\@. The variable
$x_{i,j}$ is intended to state that $i$ is mapped to $j$.  The set
$PHP_n$ consists of the following clauses:
\begin{enumerate}[$\bullet$]
\item the \emph{pigeon clause} $P_i = \bigl\{ x_{i,j} \, ;\, 1\leq j \leq n
  \bigr\}$ for every $1\leq i\leq n+1$.
\item the \emph{hole clause} $H_{i,j,k} = \{ \bar{x}_{i,k} , \bar{x}_{j,k} \}$
  for every $1\leq i<j \leq n+1$ and $k \leq n$.
\end{enumerate}

It is well-known that the pigeonhole principle requires exponential
size dag-like resolution proofs: Haken \cite{Haken85} shows that every
RD refutation of $PHP_n$ is of size $2^{\Omega(n)}$. 
Note that the number of variables is $O(n^2)$, so that this lower
bound is far from maximal. In fact, Iwama and Miyazaki \cite{iwamiy99}
prove a larger lower bound for tree-like refutations.
\begin{thm}[Iwama and Miyazaki \cite{iwamiy99}] \label{the:treelb}
  Every resolution tree refutation of $PHP_n$ is of size at least
  $(n/4)^{n/4}$.
\end{thm}
We will show that for $k\leq n/2$, RTLW($k$) refutations of $PHP_n$
are asymptotically of the same size $2^{\Omega(n\log n)}$ as
resolution trees.  On the other hand, it is known \cite{BusPit97} that
dag-like resolution proofs need not be much larger than Haken's lower
bound: there exist RD refutations of $PHP_n$ of size $2^n\cdot n^2$.
These refutations are even regular, and thus can be simulated by
regWRTI. Hence $PHP_n$ can be solved in time $2^{O(n)}$ by some
variant of \textsc{DLL-L-UP} when learning arbitrary long clauses,
whereas our lower bound shows that any DLL algorithm that learns only
clauses of size at most $n/2$ needs time $2^{\Omega(n\log n)}$.

In fact, we will prove our lower bound for the weaker
\emph{functional} pigeonhole principle $FPHP_n$, which also includes
the following clauses:
\begin{enumerate}[$\bullet$]
\item The functional clause $F_{i,j,k} = \{ \bar{x}_{i,j} , \bar{x}_{i,k} \}$
for every $1\leq i \leq n+1$ and every $1\leq j<k\leq n$. 
\end{enumerate}
While the lower bound of Iwama and Miyazaki is only stated for the
clauses $PHP_n$, it is easily verified that their proof works as well
when the functional clauses are added to the formula.

Our lower bound proof uses the fact that resolution trees with
weakening (RTW) are natural, i.e., preserved under restrictions in the
following sense:
\begin{prop}
  Let $R$ be a RTW proof of $C$ from $F$ of size $s$, and $\rho$ a
  restriction. There is an RTW proof $R'$ for $\rest{C}{\rho}$ from
  $\rest{F}{\rho}$ of size at most $s$.
\end{prop}
We denote the resolution tree $R'$ by $\rest{R}{\rho}$. Since this
proposition is well-known a proof will not be given.

Next, we need to bring refutations in RTLW($k$) to a certain normal
form.  First, we show that it is unnecessary to use clauses as lemmas
that are subsumed by axioms in the refuted formula.
\begin{lem} \label{lem:subs} If there is a RTLW($k$) refutation of
  some formula $F$ of size $s$, then there is a RTLW($k$) refutation
  of $F$ of size at most $2s$ in which no clause $C$ with $C \supseteq
  D$ for some clause $D$ in $F$ is used as a lemma.
\end{lem}
\proof
  If a clause $C$ with $C\supseteq D$ for some $D\in F$ is used as a
  lemma, replace every leaf labeled $C$ by a weakening inference of
  $C$ from $D$.
\qed
Secondly, we need the fact that an RTLW($k$) refutation does not 
need to use any tautological clauses, i.e., clauses of the form 
$C \cup \{ x , \bar{x}\}$ for a variable $x$. 
\begin{lem} \label{lem:taut} 
  If there is a RTLW($k$) refutation of some formula $F$ of size $s$,
  then there is a RTLW$(k$) refutation of $F$ of size at most $s$ that
  contains no tautological clause.
\end{lem}
\proof
  Let $P$ be an RTLW($k$)-refutation of $F$ of size $s$ that contains
  $t$ occurrences of tautological clauses. We transform $P$ into a
  refutation $P'$ of size $|P'|\leq s$ such that $P'$ contains fewer
  than $t$ occurrences of tautological clauses. Finitely many
  iterations of this process yields the claim.

  We obtain $P'$ as follows.
Since the final clause of~$P$ is not tautological, if
$t>0$, there must be a
tautological clause $C \cup
  \{x,\bar{x}\}$ which is resolved with a clause $D\cup \{x\}$
to yield a non-tautological clause $C\cup D\cup \{x\}$.
The idea is to cut out the subtree~$T_0$ that derives
the clause $C\cup\{x,\bar x\}$, and derive $C\cup D\cup\{x\}$
by a weakening from $D\cup \{ x\}$.  This gives a ``proof''~$P_0$
with fewer tautological clauses than~$P$.
  However, $P_0$~may not be a valid proof, since
  some of the clauses in~$T_0$ might be used as lemmas in $P_0$. 
To fix this, we shall extract
  parts of~$T_0$ and plant them onto~$P_0$
  so that all lemmas used are derived.  In order to make this
  construction precise, we need the notion of trees in which some of
  the used lemmas are not derived.

  A \emph {partial RTLW} from~$F$
is defined to be a tree~$T$ which satisfies all the
  conditions of an RTLW, except that some leaves may be
  labeled by clauses that occur neither in~$F$ nor earlier in
  $T$; these are called the \emph{open leaves} of~$T$.

  We construct $P'$ in stages by defining, for $i\geq 0$, a partial
  RTLW refutation~$P_i$ of~$F$ and a partial RTLW derivation~$T_i$ 
of $C\cup\{x,\bar x\}$ from~$F$
  with the following properties:
  \begin{enumerate}[$\bullet$]
  \item All open leaves in $P_i$ appear
    in $T_i$.  The first open leaf in~$P_i$ is denoted $C_i$.
  \item All open leaves in $T_i$ appear in $P_i$ before~$C_i$.
  \item $|P_i| + |T_i| = |P|$ . 
  \end{enumerate}
$P_0$ and~$T_0$ were defined above and certainly satisfy the two properties.
  Given $P_i$ and~$T_i$, we construct $P_{i+1}$ and~$T_{i+1}$ as follows:
  We locate the first occurrence of $C_i$ in $T_i$ and
  let $T^\ast_i$ be the subtree of~$T_i$ rooted at this occurrence.
  We form $T_{i+1}$ by replacing in~$T_i$ the subtree~$T^\ast_i$ by
  a leaf labeled~$C_i$.  And, we form $P_{i+1}$ by replacing the
first open leaf,~$C_i$, in~$P_i$ by the tree~$T^\ast_i$.

  The invariants are easily seen to be preserved. Obviously,
  $|P_{i+1}| + |T_{i+1}| = |P_i| + |T_i| = |P|$.  
The open leaves of~$T^\ast_i$ appear in~$P_i$ before~$C_i$, and therefore,
any open leaf in~$P_{i+1}$, and in particular,
$C_{i+1}$ if it exists, must occur after the (formerly open leaf)
clause~$C_i$.
New open
  leaves in~$T_i$ are~$C_i$ and possibly some lemmas derived in~$T^\ast_i$, 
and these all occur in~$P_{i+1}$ before~$C_{i+1}$.
  
  Since $P_{i+1}$ contains fewer open leaves than~$P_i$ for every~$i$,
there is an~$m$ such that $P_m$ contains no open leaves, and thus
  is an RTLW refutation. We then discard~$T_m$ and set $P' := P_m$.
  Each lemma used in~$P'$ was a lemma in~$P$, thus $P'$ is also an
  RTLW($k$) refutation. 

  Note that the total number of occurrences
of tautological clauses in $P_{i+1}$ and~$T_{i+1}$ combined is
the same as in $P_i$ and~$T_i$ combined. This is also equal to the
number of tautological clauses in~$P$.  Furthermore, $T_m$~must
contain at least one tautological clause, namely its root
$C\cup\{x,\bar x\}$.  It follows that $P^\prime$ has fewer tautological
clauses than~$P$.
\qed

A  matching $\rho$ is a set of pairs
$\bigl\{ (i_1,j_1) , \ldots , (i_k,j_k) \bigr\} \subset \{1,\ldots,n+1\} \times
\{ 1 ,\ldots,n\}$
such that all the $i_\nu$ as well as all the $j_\nu$ are
pairwise distinct. The size of $\rho$ is $|\rho| = k$.
A matching $\rho$ induces a partial assignment to the variables of
$PHP_n$ as follows:
\[ \rho(x_{i,j}) = \begin{cases} 1 & \text{if } (i,j) \in \rho \\
                                 0 & \text{if there is } (i,j') \in \rho \text{ with } j\neq j' \\
                                   & \text{ or } (i',j) \in \rho \text{ with } i\neq i'\\
                                 \text{undefined} & \text{otherwise.}
                   \end{cases} \] 
We will identify a matching and the assignment it induces.  
The crucial property of such a matching restriction $\rho$ is that
$\rest{FPHP_n}\rho$ is -- up to renaming of variables -- the same as
$FPHP_{n-|\rho|}$.

The next lemma states that a short clause occurring as a lemma
in an RTLW refutation can always be falsified by a small matching
restriction.
\begin{lem}\label{lem:smallrestr}
  Let $C$ be a clause of size $k \leq n/2$ such that
  \begin{enumerate}[$\bullet$]
  \item $C$ is not tautological,
  \item $C \not\supseteq H_{i,i',j}$ for any hole clause $H_{i,i',j }$,
  \item $C \not\supseteq F_{i,j,j'}$ for any functional clause $F_{i,j,j'}$.
  \end{enumerate}
  Then there is a matching $\rho$ of size $|\rho| \leq k$ such that 
  $\rest{C}{\rho} = \Box$. 
\end{lem}
\proof
  First, we let $\rho_1$ consist of all those pairs $(i,j)$ such that
  the negative literal $\bar{x}_{i,j}$ occurs in $C$.  By the second
  and third assumption, these pairs form a matching.  All the negative
  literals in $C$ are set to $0$ by $\rho_1$, and by the first
  assumption, no positive literal in $C$ is set to $1$ by $\rho_1$.

  Now consider all pigeons $i_1, \ldots , i_r$ mentioned in
  positive literals in $C$ that are not already set to $0$ by
  $\rho_1$, i.e., that are not mentioned in any of the negative
  literals in $C$.
  Pick $j_1, \ldots , j_r$ from the $n/2$ holes not mentioned in $C$,
  and set $\rho_2 := \bigl\{ (i_1 , j_1) , \ldots , (i_r,j_r) \bigr\}$.
  This matching sets the remaining positive literals to $0$, thus 
  for $\rho := \rho_1 \cup \rho_2$, we have $\rest{C}{\rho} = \Box$. 
  Clearly the size of~$\rho$ is at most~$k$ since we have picked 
  at most one pair for each literal in~$C$. 
\qed

Finally, we are ready to put all ingredients together to prove our
lower bound.
\begin{thm} \label{the:wrtlklb}
  For every $k\leq n/2$, every RTLW($k$)-refutation of $FPHP_n$
  is of size $2^{\Omega(n \log n)}$.
\end{thm}

\proof
  Let $R$ be an RTLW($k$)-refutation of $FPHP_n$ of size~$s$.  By
Lemmas \ref{lem:subs} and~\ref{lem:taut},
  $R$~can be transformed into~$R'$ of size at most~$2s$
in which no clause is tautological and
no clause used as a lemma is subsumed by a clause in $FPHP_n$.
  Let $C$ be the first clause in~$R'$ which is used as a lemma;
  $C$~is of size at most~$k$. The subtree~$R_C$ of~$R'$
rooted at~$C$ is a resolution tree for~$C$ from $FPHP_n$.

  By \pref{lem:smallrestr},
  there is a matching restriction~$\rho$ of size $|\rho|\leq k$ such
  that $\rest{C}{\rho} = \Box$.  Then $\rest{R_C}{\rho}$ is a
  resolution tree with weakening refutation of $\rest{FPHP_n}{\rho}$,
  which is the same as $FPHP_{n-k}$. By \pref{pro:rtweak},
  applications of the weakening rule can be eliminated from
  $\rest{R_C}\rho$ without increasing the size.
  Therefore by \pref{the:treelb}, $R_C$ is of size
  \[ \Bigl(\frac{n-k}{4}\Bigr)^{\frac{n-k}{4}} \geq \Bigl(\frac{n}{8}\Bigr)^{\frac{n}{8}} \] 
  and hence the size of $R$ is at least  
  $$ s \geq \frac12|R_C| \geq 2^{\Omega(n \log n)}.\eqno{\qEd}$$

%%% Local Variables: 
%%% mode: latex
%%% TeX-master: "../main"
%%% End: 

%\input{section/conclusion.tex}

\bibliography{lit.bib}
\bibliographystyle{alpha}

%% start the paper here:

\end{document}